\documentclass[preprint,12pt,3p]{elsarticle}

\usepackage{amssymb}
\usepackage{amsmath}
\usepackage{times}
\usepackage{units}
\usepackage{fullpage}
\usepackage{graphicx}
\usepackage[colorlinks=true, linkcolor=blue, citecolor=blue]{hyperref}
\usepackage{fancyhdr}
\usepackage{multirow}
\usepackage{upgreek}
\usepackage{epstopdf}
\usepackage{color,soul}
\usepackage{caption}
\usepackage{subcaption}
\usepackage{float}

\newcommand{\suma}[3]{\displaystyle\sum\limits_{#1}^{#2}{#3}}	
\newcommand{\M}[1]{\mathbf{#1}}         
\newcommand{\trn}[1]{{#1}^{\mathrm{T}}} 
\newcommand{\rve}{\Omega}               %
\newcommand{\rth}[1]{{#1}^{(r)}}        
\newcommand{\oth}[1]{{#1}^{(0)}}        
\newcommand{\sth}[1]{{#1}^{(d)}}        
\newcommand{\dev}{{\mathrm{D}}}         
\newcommand{\vol}{{\mathrm{V}}}         
\newcommand{\eps}{\varepsilon}          
\newcommand{\boldSigma}{\boldsymbol{\sigma}}    
\newcommand{\boldEps}{\boldsymbol{\varepsilon}} 
\newcommand{\subs}[1]{_{\mathrm{#1}}}   
\newcommand{\lev}[1]{^{\mathrm{#1}}}    

\newcommand{\etal}{~et~al.}

\hypersetup{
    pdfauthor = {Vaclav Nezerka},
    pdftitle = {Micromechanical Model to Predict Tensile Properties of Lime-Based Mortars},
    pdfsubject = {Modeling of Mortars},
    pdfkeywords = {micromechanics, tensile properties, Mori-Tanaka, lime-based mortars, shrinkage-induced cracking}
}

\journal{arXiv}

\begin{document}

\begin{frontmatter}



\author[ctu]{V. Ne\v{z}erka\corref{cor1}}
\ead{vaclav.nezerka@fsv.cvut.cz}
\cortext[cor1]{Corresponding author}
\author[ctu]{J. N\v{e}me\v{c}ek}
\ead{jiri.nemecek@fsv.cvut.cz}
\author[ctu]{J. Zeman}
\ead{zemanj@cml.fsv.cvut.cz}

\address[ctu]{Faculty of Civil Engineering, Czech Technical University in Prague, Th\'{a}kurova~7, 166~29 Praha~6, Czech Republic}

\title{Micromechanics-Based Simulations of Compressive and Tensile Testing on Lime-Based Mortars}

\begin{abstract}
The purpose of this paper is to propose a continuum micromechanics model for the simulation of uniaxial compressive and tensile tests on lime-based mortars, in order to predict their stiffness, compressive and tensile strengths, and tensile fracture energy. In tension, we adopt an incremental strain-controlled form of the Mori-Tanaka scheme with a damageable matrix phase, while a simple $J_2$ yield criterion is employed in compression. To reproduce the behavior of lime-based mortars correctly, the scheme must take into account shrinkage cracking among aggregates. This phenomenon is introduced into the model via penny-shaped cracks, whose density is estimated on the basis of a particle size distribution combined with the results of finite element analyses of a single crack formation between two spherical inclusions. Our predictions show a good agreement with experimental data and explain the advantages of compliant crushed brick fragments, often encountered in ancient mortars, over stiff sand particles. The validated model provides a reliable tool for optimizing the composition of modern lime-based mortars with applications in conservation and restoration of architectural heritage.
\end{abstract}

\end{frontmatter}


\section{Introduction}

The lime-based mortars were widely used as masonry binder in ancient times~\cite{Moropoulou_2000, Maravelaki_2003}. Nowadays, they are often required by the authorities for cultural heritage for repairs of old masonry because of their compatibility with the original materials~\cite{Sepulcre_2010, Callebaut_2001}. The substitution of lime-based mortars with binders based on Portland cement turned out to be inappropriate, because of the damage to the original masonry due to high stiffness contrast and presence of solutable salts~\cite{Veniale_2003}. On the other hand, the calcitic matrix of pure lime mortars is relatively weak, more compliant~\cite{Arizzi_2012, Lanas_2004}, and susceptible to shrinkage up to 13~\%~\cite{Nezerka_2013_pastes, Wilk_2013}.

To improve the durability and strength of ancient mortars, masons often used additives rich in silica (SiO$_2$) and alumina (Al$_2$O$_3$)~\cite{Arizzi_2012, Velosa_2009, Papayianni_2007} in the form of volcanic ash or crushed ceramic bricks, tiles or pottery~\cite{Moropoulou_2005, Farci_2005}. In recent years, metakaolin has become a very popular alternative to these ancient additives, because of its high reactivity~\cite{Arizzi_2012, Nezerka_2013_pastes, Velosa_2009}. On the other hand, the crushed brick fragments are considered rather as an inert aggregate since the hydraulic reaction, if any, can take place only at the interface between the fragments and surrounding matrix. Moreover, the formation of hydraulic products requires the presence of moisture~\cite{Baronio_1997_2}, significant amount of time~\cite{Moropoulou_2002}, and ceramic clay fired at appropriate temperatures~\cite{Bakolas_2008}. Beside the matrix-enhancing additives, the mechanical properties of lime mortars can be also improved by optimizing the amount and composition of aggregates~\cite{Arizzi_2012, Lanas_2004, Stefanidou_2005}, mostly via a time-demanding trial-and-error procedure.

The goal of this paper is to render the design process more efficient by proposing a simple model for the prediction of basic properties of lime-based mortars in tension and compression. Our developments have been inspired by earlier studies dealing with micromechanics of cement pastes in compression~\cite{Pichler_2011} and in tension~\cite{Vorel_2012}. The first study by Pichler and Hellmich~\cite{Pichler_2011} exploits a two-level homogenization approach combining the self-consistent~\cite{Hill_1965_2, Budiansky_1965} and Mori-Tanaka~\cite{Mori_1973, Benveniste_1987} schemes and a $J_2$-based criterion to estimate the compressive strength. In the latter work, Vorel\etal~\cite{Vorel_2012} estimated the tensile strength and fracture energy by combining the incremental form of the Mori-Tanaka method at a single-level with the crack band model~\cite{Bazant_1983} to account for the distributed matrix cracking. However, the crucial feature missing in both models is the effect of shrinkage-induced cracks that are intrinsic to the mechanical properties of lime-based mortars.

The shrinkage-induced cracking in cement-like materials has been addressed by both modelling and experimental studies. Detailed analytical investigation into shrinkage cracking around a single cylindrical aggregate was performed by Dela and Stang~\cite{Dela_2000} to estimate crack growth in high-shrinkage cement paste. Behavior of multi-aggregate systems was addressed numerically by~Grassl and Wong~\cite{Grassl_2010} using a discrete lattice model; their findings were in agreement with the cracking patterns observed by Bisschop and Mier~\cite{Bisschop_2002}. Backscattered electron microscopy (BSE) confirmed that the lime-based mortars rich in sand suffer from an extensive matrix cracking~\cite{Stefanidou_2005}. Based on these studies and also on our independent experimental investigations~\cite{Nezerka_2013_pastes, Nezerka_2014_nanoindentation}, we decided to represent the shrinkage cracks as penny shaped polydisperse voids in our homgenization scheme.

Based on these considerations, the model proposed in Section~\ref{sec:model} operates at two scales, see Figure~\ref{fig:scheme}. At Level~I, we account for the individual components of mortar, such as lime matrix, sand or brick particles, and distributed voids. At Level~II, the shrinkage cracks are introduced into the homogenized material from Level~I. Details of this procedure are provided in Section~\ref{sec:elasticity}, with the goal to estimate initial elastic properties by the Mori-Tanaka procedure at Level~I, Section~\ref{sec:elasticity-levelI}, and the dilute approximation at Level~II, Section~\ref{sec:elasticity-levelII}. The density and size distribution of the penny-shaped shrinkage cracks are determined from a crack formation criteria, proposed in Section~\ref{sec:crackDensityParameter} on the basis of three-dimensional finite element analyses of shrinkage-induced cracking between two isolated inclusions. Two extensions of the elastic model are presented next. The strength under stress-controlled uniaxial compression is estimated in Section~\ref{sec:compressiveStrengthEstimation} on the basis of the $J_2$ stress invariant in the matrix phase. Under strain-controlled uniaxial tension, Section~\ref{sec:incremental_MT_tension}, we employ the incremental form of the Mori-Tanaka scheme coupled with an isotropic damage constitutive model to estimate the tensile strength and fracture energy.

Having introduced our model, in Section~\ref{sec:experiments} we specify the input data for individual components, along with the experimental procedures used to acquire them, and the composition of the tested mortar samples. Section~\ref{sec:results} is dedicated mostly to the model validation, concluded by the determination of the optimal mix composition. Finally, we summarize our results in Section~\ref{sec:conclusions} and outline the strategy how to translate them to the structural scale.

In the following text, the condensed Mandel representation of symmetric tensorial quantities is employed, e.g.~\cite{Milton_2002}. In particular, the scalar quantities are written in the italic font, e.g.~$a$ or $A$, and the boldface font, e.g.~$\M{a}$ or $\M{A}$, is used for vectors or matrices representing second- or fourth-order tensors. $\trn{\M{A}}$ and $(\M{A})^{-1}$ denote the matrix transpose and the inverse matrix, respectively. Other symbols are introduced later, when needed.

\section{Model} \label{sec:model}
\begin{figure}[ht]
\centering
    \def\svgwidth{0.9\linewidth}
    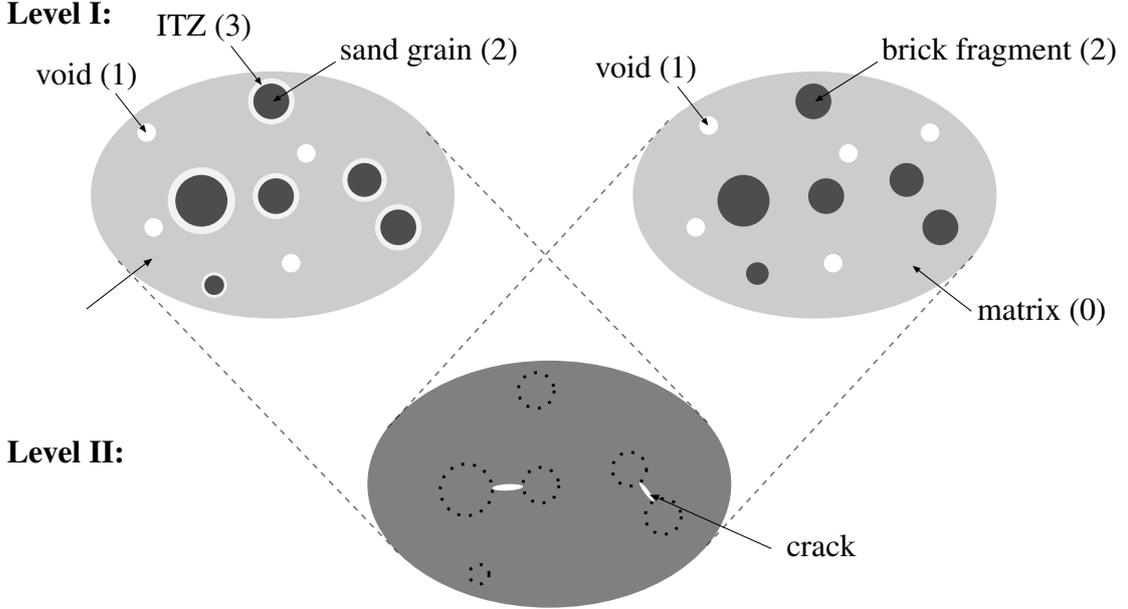
\caption{Scheme of the micromechanical model of mortars with various aggregate-types; the numbers in parentheses refer to the indexes of individual phases.} \label{fig:scheme}
\end{figure}

We consider an RVE occupying a domain~$\rve$, composed of $m$ phases indexed by $r$ at Level~I, and penny-shaped shrinkage-induced cracks reflected at Level~II. The matrix is represented by $r = 0$ and indexes $r = 1,...,m$ refer to heterogeneities of spherical shape or spherical shell in the case of interfacial transition zone (ITZ) around sand grains, see Figure~\ref{fig:scheme}. The volume fraction of $r$-th phase, having the volume $|\rth{\rve}|$, is provided by $\rth{c} = |\rth{\rve}|/|\rve|$. Note the representation of sand / crushed brick particles by spheres is less realistic then e.g. by ellipsoids~\cite{Pichler_2009}, but the effect of the introduced errors is minor relative to the accuracy of the input data, cf. Section~\ref{sec:experiments}.

\subsection{Elasticity}\label{sec:elasticity}

\subsubsection{Level I: Homogenization of Aggregates and Voids} \label{sec:elasticity-levelI}
The elastic response of individual phases is described by the material stiffness matrix $\rth{\M{L}}$. Since all phases are considered as geometrically and materially isotropic, the matrix~$\rth{\M{L}}$ can be decomposed using the orthogonal volumetric and deviatoric projections $\M{I}_\vol$ and $\M{I}_\dev$, e.g.~\cite[p.~23]{Milton_2002},
\begin{equation} \label{eq:decomposition}
   \rth{\M{L}} = \rth{3K}\M{I}_\vol + \rth{2G}\M{I}_\dev,
\end{equation}
where $\rth{K}$ and $\rth{G}$ denote the bulk and shear moduli of the $r$-th phase.

Under the dilute approximation, the mean strain in individual phases, $\rth{\boldEps}$, is related to the macroscopic strain, $\boldEps$, via the dilute concentration factors,
\begin{equation} \label{eq:Adil}
   \rth{\boldEps} = \rth{\M{A}}\subs{dil} \boldEps.
\end{equation}
In the Mori-Tanaka scheme, the strain in individual phases can be found as $\rth{\boldEps} = \rth{\M{A}}\subs{dil} \oth{\boldEps}$, where $\oth{\boldEps}$ is the strain within the matrix found as
\begin{equation} \label{eq:AMT}
   \oth{\boldEps} = \M{A}\subs{MT} \boldEps,
\end{equation}
where the strain concentration factor $\M{A}\subs{MT}$ is provided by
\begin{equation} \label{eq:AMT_formula}
   \M{A}\subs{MT} = \biggl(\oth{c}\M{I} + \suma{r=1}{m}{\rth{c}\rth{\M{A}}\subs{dil}} \biggr)^{-1}.
\end{equation}

Because of isotropy, the effective stiffness at Level I is fully specified by the effective bulk, $K\lev{I}\subs{eff}$, and shear, $G\lev{I}\subs{eff}$, moduli
\begin{align} \label{eq:Keff_and_Geff}
	K\lev{I}\subs{eff} = \cfrac
{\oth{c} \oth{K} + \suma{r=1}{m}{\rth{c} \rth{K} \rth{A}\subs{dil,V}}}
{\oth{c} + \suma{r=1}{m}{\rth{c} \rth{A}\subs{dil,V}}},
&& 
    G\lev{I}\subs{eff} = \cfrac
{\oth{c} \oth{G} + \suma{r=1}{m}{\rth{c} \rth{G} \rth{A}\subs{dil,D}}}
{\oth{c} + \suma{r=1}{m}{\rth{c} \rth{A}\subs{dil,D}}}
\end{align}
that depend on the volumetric and deviatoric components of the dilute concentration factors
\begin{equation}\label{eq:A_decomposition}
	\rth{\M{A}}\subs{dil} = \rth{A}\subs{dil,V} \M{I}_\vol + \rth{A}\subs{dil,D} \M{I}_\dev, r=1,...,m.
\end{equation}

\paragraph{Uncoated Inclusions}
The dilute concentration factors for spherical particles follow from the seminal Eshelby work~\cite{Eshelby_1957}:
\begin{align}\label{eq:AdilV_AdilD}
	\rth{A}\subs{dil,V} = \cfrac{\oth{K}}{\oth{K}+\oth{\alpha}(\rth{K}-\oth{K})}, &&
    \rth{A}\subs{dil,D} = \cfrac{\oth{G}}{\oth{G}+\oth{\beta}(\rth{G}-\oth{G})},
\end{align}
where $\oth{\alpha}$ and $\oth{\beta}$ depend on the matrix Poisson's ratio, $\oth{\nu}$, as:
\begin{align} \label{eq:EshelbySpherical}
    \oth{\alpha} = \cfrac{1+\oth{\nu}}{3(1+\oth{\nu})},
    &&
    \oth{\beta} = \cfrac{2(4-5\oth{\nu})}{15(1-\oth{\nu})}.
\end{align}

\paragraph{Coated Inclusions}
The more involved case of particles coated by spherical shells was solved by Herv\'{e} and Zaoui~\cite{Herve_1993}. Their introduction into the scheme makes the model sensitive to the grain-size distribution because the strain concentration factors depend on the radius of sand grains relative to their coating. From that reason the sand aggregates (2) and ITZ (3), Figure~\ref{fig:scheme}, must be subdivided into sub-phases $m^{\delta}$ corresponding to individual grain-size intervals. These are denoted by indices ($2,\delta$) and ($3,\delta$), where $\delta = 1,...,m^{\delta}$.

Spatially, the dilute concentration factors for sand grains and surrounding ITZ, both represented by their outer radii, $R^{(2,\delta)}$ and $R^{(3,\delta)}$, and Poisson's ratios, $\nu^{(2)}$ and $\nu^{(3)}$, in the form
\begin{equation}\label{eq:AdilV_coated}
	A^{(2,\delta)}\subs{dil,V} = \cfrac{1}{Q_{11}^2}, \quad
    A^{(3,\delta)}\subs{dil,V} = \cfrac{Q_{11}^1}{Q_{11}^2}
\end{equation}
and
\begin{equation}\label{eq:AdilD_coated}
    A^{(2,\delta)}\subs{dil,D} = A_1 - \cfrac{21}{5} \: \cfrac{{R^{(2,\delta)}}^2}{1-2\nu^{(2)}} B_1, \quad
    A^{(3,\delta)}\subs{dil,D} = A_2 - \cfrac{21}{5} \: \cfrac{{R^{(3,\delta)}}^5-{R^{(2,\delta)}}^5}{(1-2\nu^{(3)})({R^{(3,\delta)}}^3-{R^{(2,\delta)}}^3)} B_2,
\end{equation}
where the auxiliary factors $Q_{11}^1$, $Q_{11}^2$, $A_1$, $A_2$, $B_1$ and $B_2$ are provided in~\cite[Appendix A]{Nezerka_2012_AP}.

The volume fractions of the coatings, $\{c^{(3,\delta)}\}_{\delta=1}^{m^{\delta}}$, are determined from the given volume fractions of the sand sub-phases, $\{c^{(3,\delta)}\}_{\delta=1}^{m^{\delta}}$, as
\begin{equation}
	c^{(3,\delta)} = \biggl (\biggl(\cfrac{R^{(3,\delta)}}{R^{(2,\delta)}}\biggr)^3-1 \biggr )c^{(2,\delta)}.
\end{equation}
Because the coatings are assume to replace the matrix surrounding the inclusions, their total volume fraction $\suma{\delta=1}{m^{\delta}}{c^{(3,\delta)}}$ is subtracted from the matrix volume fraction $\oth{c}$.

\subsubsection{Level II: Homogenization of Shrinkage-Induced Cracks}\label{sec:elasticity-levelII}
The effective stiffness at Level II accounts for the effects of randomly distributed cracks under the dilute approximation. This reduction is accomplished by introducing additional compliances $H\subs{E}$ and $H\subs{G}$~\cite{Kachanov_2005}:
\begin{align}\label{eq:crackCompliance}
    H\subs{E} = \cfrac{f}{E\lev{I}\subs{eff}} \: \cfrac{16(1-{\nu\lev{I}\subs{eff}}^2)(10-3\nu\lev{I}\subs{eff})}{45(2-\nu\lev{I}\subs{eff})},
&&
    H\subs{G} = \cfrac{f}{E\lev{I}\subs{eff}} \: \cfrac{32(1-{\nu\lev{I}\subs{eff}}^2)(5-\nu\lev{I}\subs{eff})}{45(2-\nu\lev{I}\subs{eff})},
\end{align}
where $f$ is the crack density parameter, determined later in Section~\ref{sec:crackDensityParameter}, $E\lev{I}\subs{eff}$ and $\nu\lev{I}\subs{eff}$ are the effective Young's modulus and the Poisson ratio at Level~I, obtained from~\eqref{eq:Keff_and_Geff} through, e.g.~\cite[p.~23]{Milton_2002},
\begin{align}
E\lev{I}\subs{eff} = \cfrac{9K\lev{I}\subs{eff}G\lev{I}\subs{eff}}{3K\lev{I}\subs{eff}+G\lev{I}\subs{eff}},
&&
\nu\lev{I}\subs{eff} = \cfrac{3K\lev{I}\subs{eff}-2G\lev{I}\subs{eff}}{2(3K\lev{I}\subs{eff}+G\lev{I}\subs{eff})},
\end{align}
The effective Young's and shear moduli of the cracked composite then follow from
\begin{align}\label{eq:EeffCr}
E\lev{II} = \biggl(\cfrac{1}{E\lev{I}\subs{eff}}+H\subs{E} \biggr)^{-1}, &&
G\lev{II}\subs{eff} = \biggl(\cfrac{2(1+\nu\lev{I}\subs{eff})}{E\lev{I}\subs{eff}}+H\subs{G} \biggr)^{-1}.
\end{align}

\subsubsection{Crack Density Estimation}\label{sec:crackDensityEstimation}

The criteria for the formation of shrinkage cracks were established based on results of 3D FE model containing 200k to 700k tetrahedral linear elements, depending on the RVE size. The average average element size was equal to~0.1~mm. The RVE consisted of two spherical inclusions embedded in a matrix. The model was supported in three corners in such a way to allow contraction of the RVE in all directions without external constraints. The shrinkage was introduced into the model via incrementally increased matrix eigenstrain and the resulting system was solved by the Newton-Raphson algorithm. Because the study was focused on establishing the critical gap between the inclusions, the simulations were stopped after the crack between the inclusions appeared or if the eigenstrain reached $4~\%$, since such value corresponds to the maximum shrinkage of lime-based pastes~\cite{Nezerka_2013_pastes}.

\begin{figure}[htp]
\centering
   \begin{subfigure}{0.4\linewidth} \centering
     \includegraphics[width=\textwidth]{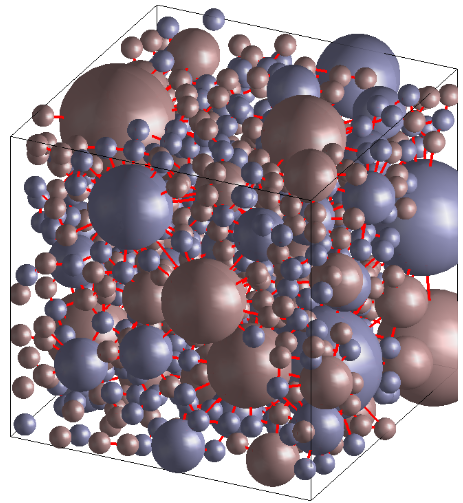}
     \caption{example of polydisperse system RVE simulation ($c^{(2)}=0.2$; the red lines indicate diameters of cracks between aggregates (if formed)}
   \end{subfigure}
   \hspace{0.05\linewidth}
   \begin{subfigure}{0.46\linewidth} \centering
     \includegraphics[width=\textwidth]{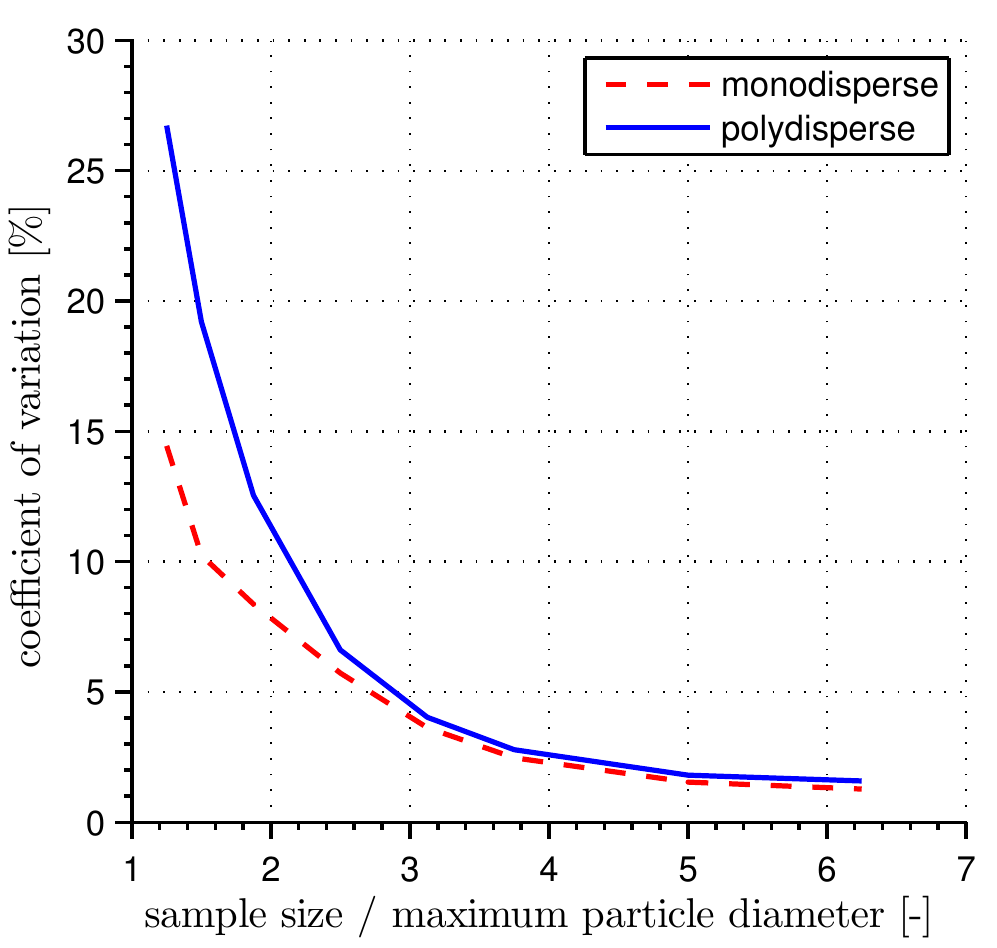}
     \caption{relationship between RVE size relative to the maximum particle diameter and variation of the crack density parameters (based on 200~simulations)}
     \label{fig:convergence}
   \end{subfigure}
\caption{MATLAB simulations of shrinkage-induced cracking within RVE}
\label{fig:crackFormationRVE}
\end{figure}

The simulations were performed in OOFEM software~\cite{Patzak_2001}, utilizing the anisotropic damage material model by Jir\'{a}sek~\cite{Jirasek_1999} for the matrix, and the isotropic elastic model for the aggregates. The anisotropic damage material model utilizes the concepts from the microplane theory --- the damage variable characterizes the relative compliance of the material for each microplane direction and therefore the stiffness is not reduced parallel to the crack. The material parameters of individual phases used for the analysis are summarized in Table~\ref{tab:materialProperties}. ITZ around aggregates was not explicitly modeled; however, the zone of damage around aggregates developed spontaneously due to tensile stresses perpendicular to the grains at the interface, Figure~\ref{fig:crackFormationRVE}). The mesh-independence was ensured by the crack band approach~\cite{Bazant_1983}.

\begin{figure}[htp]
\centering
   \begin{subfigure}{0.42\linewidth} \centering
     \includegraphics[width=\textwidth]{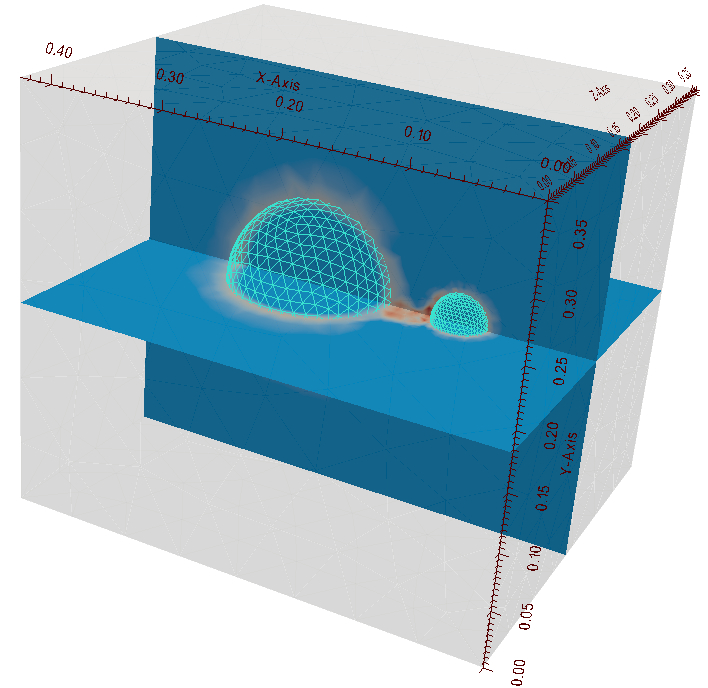}
   \end{subfigure}
   \hspace{0.1\linewidth}
   \begin{subfigure}{0.35\linewidth} \centering
     \includegraphics[width=\textwidth]{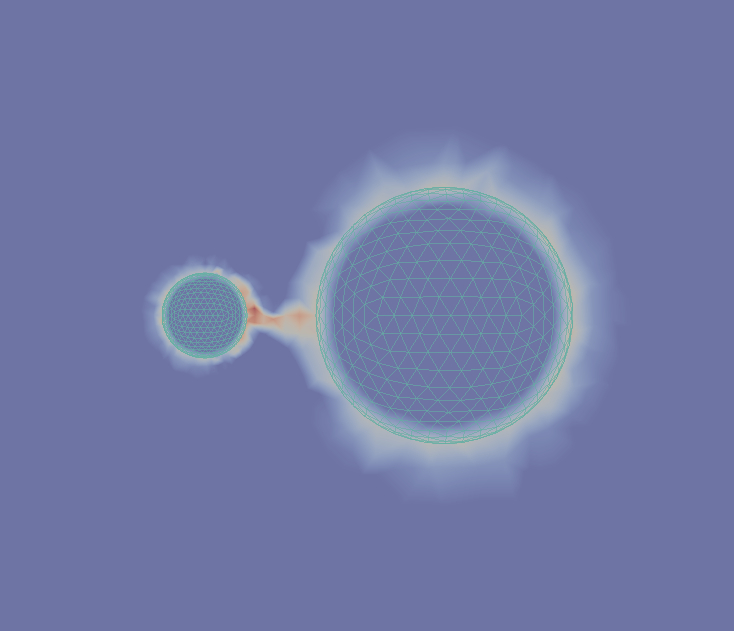}
     \vspace{13pt}
   \end{subfigure}
\caption{FE simulation of the crack formation between two aggregates of a size equal to 0.3 and 1.0~mm, respectively; visualized by ParaView software: axonometric view (left), front view (right). The red color indicates the formed cracks (the damage variable is equal to 1.0), while the blue regions represent the matrix volume with no damage.} \label{fig:crackFormation}
\end{figure}

\begin{figure}[htp]
\centering
\includegraphics[width=0.9\textwidth]{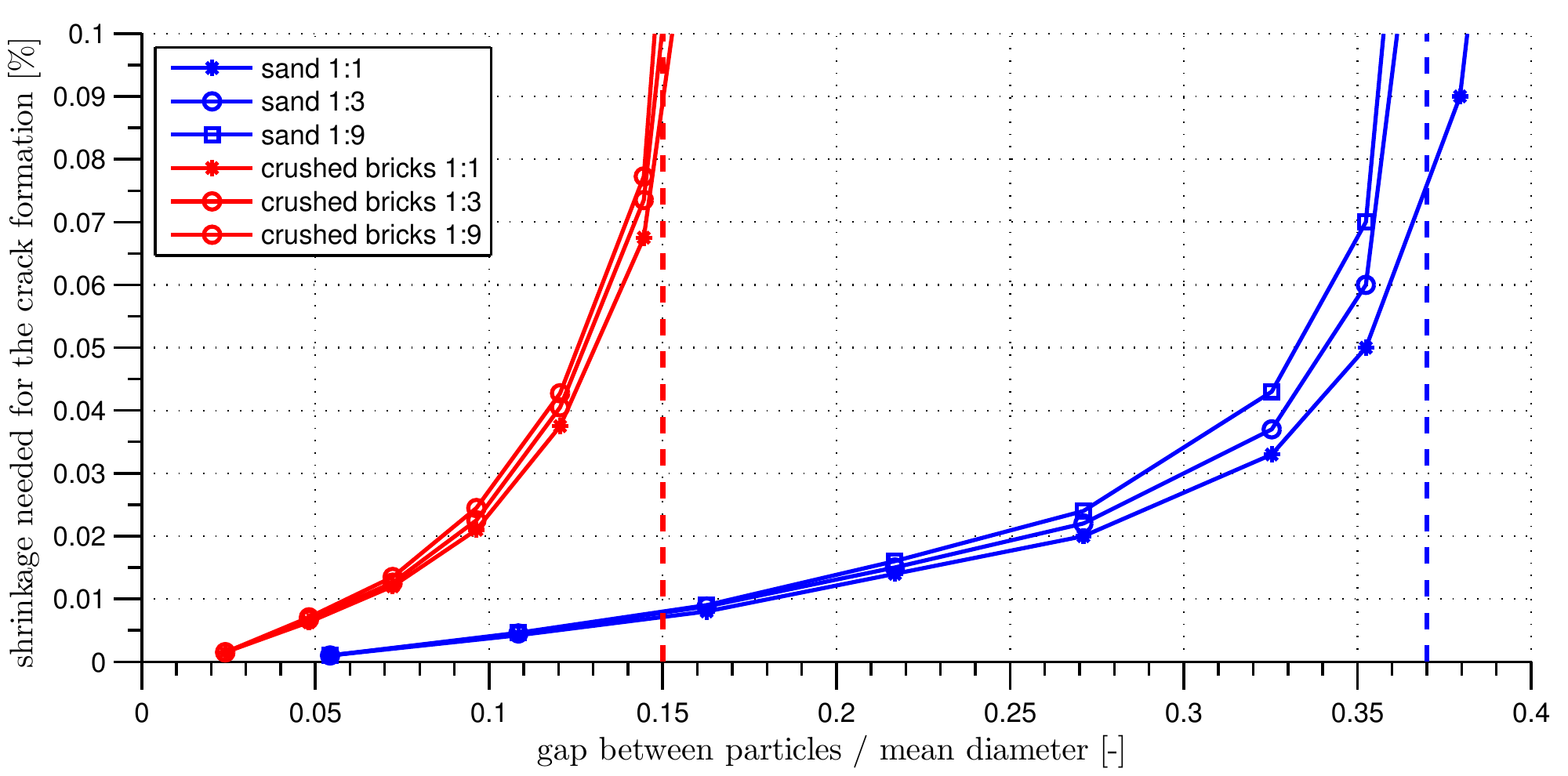}
\caption{Critical relative gap, $g\subs{crit}$, indicated by dashed line for the compliant brick and stiff sand inclusions of three ratios of their diameters, when embedded in high-shrinkage (significantly exceeding $0.1~\%$) matrix.}\label{fig:criticalRatio}
\end{figure}

The results of the simulations appear in Figure~\ref{fig:criticalRatio}, in which we plot the critical shrinkage strain as a function of the dimensionless gap between the particles,
\begin{equation} \label{eq:gap}
   g_{12} = \cfrac{2l_{12}}{d_1+d_2},
\end{equation}
where $l_{12}$ stands for the face-to-face particle distance, and $d_1$ and $d_2$ are the particle diameters. Two particle types~(compliant brick and stiff sand) and three ratios of the particle diameters were considered. For the shrinkage strain exceeding $\approx 0.1~\%$~(which is much below the typical value of $\approx 4~\%$ measured in~\cite{Nezerka_2013_pastes}), we observe that a shrinkage crack between the particles develops if the gap is smaller than a critical value $g\subs{crit}$; otherwise no crack forms. The critical value is only weakly dependent on the diameter ratio, but depends strongly on the particle stiffness -- in what follows we consider $g\subs{crit} = 0.37$ for the sand aggregates and $g\subs{crit} = 0.15$ for the brick aggregates, respectively.

\paragraph{Crack Density Parameter}\label{sec:crackDensityParameter}
A circular penny-shaped crack is assumed to form between two neighboring particles once their dimensionless gap is smaller than $g\subs{crit}$. Given the number of the particles, $k$, distributed within a represenative volume element~(RVE), $\rve$, the crack density parameter is defined as~\cite{Kachanov_2005}
\begin{equation} \label{eq:crackDensityParameter}
   f = \cfrac{1}{|\rve|}\sum_{\{i,j=1,\ldots,k: g_{ij} \leqq g\subs{crit}\}}{(l_{ij})^3},
\end{equation}
recall Eq.~\eqref{eq:gap}.

The input data to the crack density analysis were generated by an in-house packing algorithm for polydisperse spheres implemented in MATLAB. For the target volume fraction of the particles, $c^{(2)}$, and the given curve, we generate the particle distribution by a random sequential addition algorithm, proceeding from the largest particles to the smallest ones. As for the RVE size, we found that the size of 5-times maximum particle diameter yields the coefficient of variation in the density parameter $f$ less than $\approx 2~\%$, Figure~\ref{fig:convergence}, which is sufficient for practical purposes.

\subsection{Compressive Strength} \label{sec:compressiveStrengthEstimation}

Inspired by phenomenological, e.g.~\cite{Feenstra_1995}, and micromechanical, e.g.~\cite{Pichler_2011}, models for cementitious materials, we adopt the von-Mises failure criterion for the matrix phase,
\begin{equation}\label{eq:failure_condition}
\sqrt{J_2^{(0)}} - \frac{f\subs{c}^{(0)}}{\sqrt{3}} = 0,
\end{equation}
to estimate the mortar strength in compression. In Eq.~\eqref{eq:failure_condition}, $f\subs{c}^{(0)}$ denotes the matrix compressive strength, and the second deviatoric invariant in the matrix phase, $J_2^{(0)}$, is determined from the average matrix stress, $\oth{\boldSigma}$, through
\begin{equation}
J_2^{(0)} = \cfrac{1}{2} \trn{\oth{\boldSigma}} \M{I}_\dev \oth{\boldSigma}.
\end{equation}
The average stress in the matrix phase $\oth{\boldSigma}$  follows from the basic assumption of the Mori-Tanaka method, Eq.~\eqref{eq:AMT} as
\begin{equation}
\oth{\boldSigma}
=
\oth{\M{L}}
\M{A}\subs{MT}
\left(\M{L}\lev{II}\subs{eff} \right)^{-1}
\boldSigma,
\end{equation}
where $\boldSigma$ stands for the applied macroscopic stress. The mortar compressive strength $f\subs{c}$ is found by subjecting the sample to the macroscopic uniaxial stress state $\boldSigma = \trn{[-f\subs{c},0,0,0,0,0]}$, such that the condition~\eqref{eq:failure_condition} is satisfied.

\subsection{Tensile strength and Fracture energy} \label{sec:incremental_MT_tension}

Primarily the matrix phase is subjected to damage, but also the ITZ stiffness must be reduced to avoid locking effects. The materials' softening is driven by the damage evolution in the matrix phase at Level I
\begin{equation}\label{eq:stress_strain}
   \sth{\boldSigma} = (1-\oth{\omega})
   \sth{\M{L}} \sth{\boldEps}
   = \sth{\M{L}}_\mathrm{sec} \sth{\boldEps},
\end{equation}
where $d = 0, 3$ refers to the damageable phases. Because we assume that the strain in the matrix is increasing monotonically, the magnitude of damage in the matrix phase follows from
\begin{equation}\label{eq:damage_strain}
\oth{\omega} = 1- \cfrac{\oth{\eps_0}}{\oth{\eps}\subs{eq}}\exp\biggl(-\cfrac{\oth{\eps}\subs{eq}-\oth{\eps_0}}{\oth{\eps}\subs{f}-\oth{\eps_0}}\biggr),
\end{equation}
where $\oth{\eps}\subs{eq}$ denotes the Rankine effective strain, determined from the tensile parts of the principal strains $\langle \oth{\eps}_I \rangle_+$ via
\begin{equation}
\oth{\eps}\subs{eq}
=
\| \oth{\boldEps} \|
=
\max_{I=1}^{3} \langle \oth{\eps}_I \rangle_+.
\end{equation}
The critical equivalent strains at the damage onset, $\oth{\eps}\subs{0}$, and the fracturing strain, $\oth{\eps}\subs{f}$, follow from
\begin{align}
\oth{\eps}\subs{0} = \cfrac{\oth{f}\subs{t}}{\oth{E}},
&&
\oth{\eps}\subs{f} = \cfrac{\oth{G}\subs{f}}{\oth{f}\subs{t}h},
\end{align}
where $\oth{f}\subs{t}$ is the matrix tensile strength, $\oth{G}\subs{f}$ represents the matrix fracture energy, and $h$ is the crack band width of the strain-softening zone~\cite{Bazant_1983}, set to $2.7$ times the maximum aggregate~\cite{Bazant_1989}.

In order to estimate the effective tensile properties, we subject the composite to incremental strain path ($n = 1,2,\ldots$)
\begin{equation}
   \boldEps_{n} = \trn{[\eps_{1,n-1}+\Delta\eps,0,0,0,0,0]},
\end{equation}
where $\Delta\eps$ is a fixed strain increment, equal to $10^{-7}$, and $\boldEps_0 = \boldsymbol{0}$. The damage in the matrix phase at the $n$-th load step, $\oth{\omega}_n$, is determined from the consistency condition between the damage value in~\eqref{eq:stress_strain}, related to the reduction of the matrix and ITZ stiffness, and the value determined from~\eqref{eq:damage_strain} with the local strain $\oth{\boldEps}$ estimated by the elastic Mori-Tanaka method from Eq.~\eqref{eq:AMT} with adjusted phase stiffnesses $\sth{\M{L}}_\mathrm{sec}$ and the overall strain set to $\boldEps_{n}$. The resulting single non-linear equation for $\oth{\omega}_n$ is solved by the conventional secant method on the interval $[\oth{\omega}_{n-1};1]$ with the accuracy set to $10^{-8}$. The simulation is stopped when the damage variable $\oth{\omega}$ reaches the value of $0.999$; that corresponds to the $N$-th time step.

The effective tensile parameters are obtained post-processing of the history of macroscopic stresses
\begin{align}
\boldSigma_n
=
\M{L}\lev{II}_{\mathrm{eff},n}
\boldEps_n
\text{ for }
n = 1, 2, \ldots, N.
\end{align}
In particular, the effective tensile strength $f\subs{t}$ is estimated as
\begin{equation}
f\subs{t} =
\max_{n=1}^{N} \| \boldSigma_n \|,
\end{equation}
while the effective fracture energy $G\subs{f}$ follows from
\begin{equation}
G\subs{f} = \frac{h \Delta\epsilon}{2}
\sum_{n=1}^{N} \left(
\sigma_{n-1,1} + \sigma_{n,1}
\right).
\end{equation}

\section{Experimental Analysis}\label{sec:experiments}

In this section, we gather the results of experimental studies needed to acquire the input data to the model, summarized in Table~\ref{tab:materialProperties}, as well as the validation data. After introducing the mortar components, Section~\ref{sec:materials}, in Section~\ref{sec:elastic} we discuss the acquisition of the elastic properties of individual phases. Section~\ref{sec:inelastic} is dedicated to inelastic properties of the matrix phase and the validation data are introduced in Section~\ref{sec:validation}. Note that all experiments were always carried out on at least six specimens representing the same material or batch, in order to obtain representative data. In addition, when possible, we also compared our results with independent results from the literature, in order to ensure their credibility.

\begin{table}[ht]
 \caption{Material properties of individual phases; $\rho$, $E$, $\nu$, $f\subs{c}$, $f\subs{t}$, $G\subs{f}$, and PSD denote the mass density, Young's modulus, Poisson's ratio, compressive strength, tensile strength, fracture energy, and particle-size distribution, respectively.}
 \label{tab:materialProperties}
 \centering
 \renewcommand{\arraystretch}{1.2}
 \begin{tabular}{c c c c c c c c}
  \hline
  phase          & $\rho$     & $E$     & $\nu$  & $f\subs{c}$ & $f\subs{t}$ & $G\subs{f}$ & PSD       \\
                 & [kg/m$^3$] & [GPa]   & [-]    & [MPa]       & [MPa]       & [J/m$^2$]   &           \\ \hline
  matrix         & 1,066       & 8.00    & 0.25   & 7.0         & 2.0         & 12.0        & $\times$  \\
  voids          & $\times$   &    0    & 0.25   & $\times$    & $\times$    & $\times$    & no need   \\
  crushed bricks & 1,761       & 3.50    & 0.2    & $\times$    & $\times$    & $\times$    & needed    \\
  siliceous sand & 2,720       & 90.0    & 0.17   & $\times$    & $\times$    & $\times$    & needed    \\
  ITZ            & 1,066       & 2.67    & 0.25   & $\times$    & $\times$    & $\times$    & thickness of 10~$\upmu$m  \\
  \hline
 \end{tabular}
\end{table}

\subsection{Materials}\label{sec:materials}

The investigated mortars were reinforced by sand or crushed brick fragments, and the desired grain size distributions were obtained via sieving. The binder was composed of commonly available white air-slaked lime~(CL90) \v{C}ertovy schody, Czech Republic, of a high purity (98.98~\% of CaO + MgO), and metakaolin (finely ground burnt claystone, commercial name Mefisto L05) in a mass ratio equal to 7:3. The amount of water was adjusted so that the fresh mortars fulfilled the workability slump test, which is set according to \v{C}SN~EN~1015-3 as~15$\pm$1~cm.

The volume of individual phases listed in Table~\ref{tab:materialProperties} was calculated from the mass of mortar constituents and the measured mass density, $\rho$. The matrix mass was determined based on the result of our previous study~\cite{Nezerka_2013_pastes}, which revealed that~1~kg of slaked lime powder and metakaolin mixed in the ratio 7:3 produce 1.46~kg of the paste. The volume of voids was obtained experimentally using the pycnometer method. For detailed information on the conversion of mass fractions to the volume volume fractions for individual phases, the reader is referred to~\cite[Section 3]{Nezerka_2012_AP}.

\subsubsection{Particle-Size Distribution}

\begin{figure}[htp]
\centering
\includegraphics[width=0.5\textwidth]{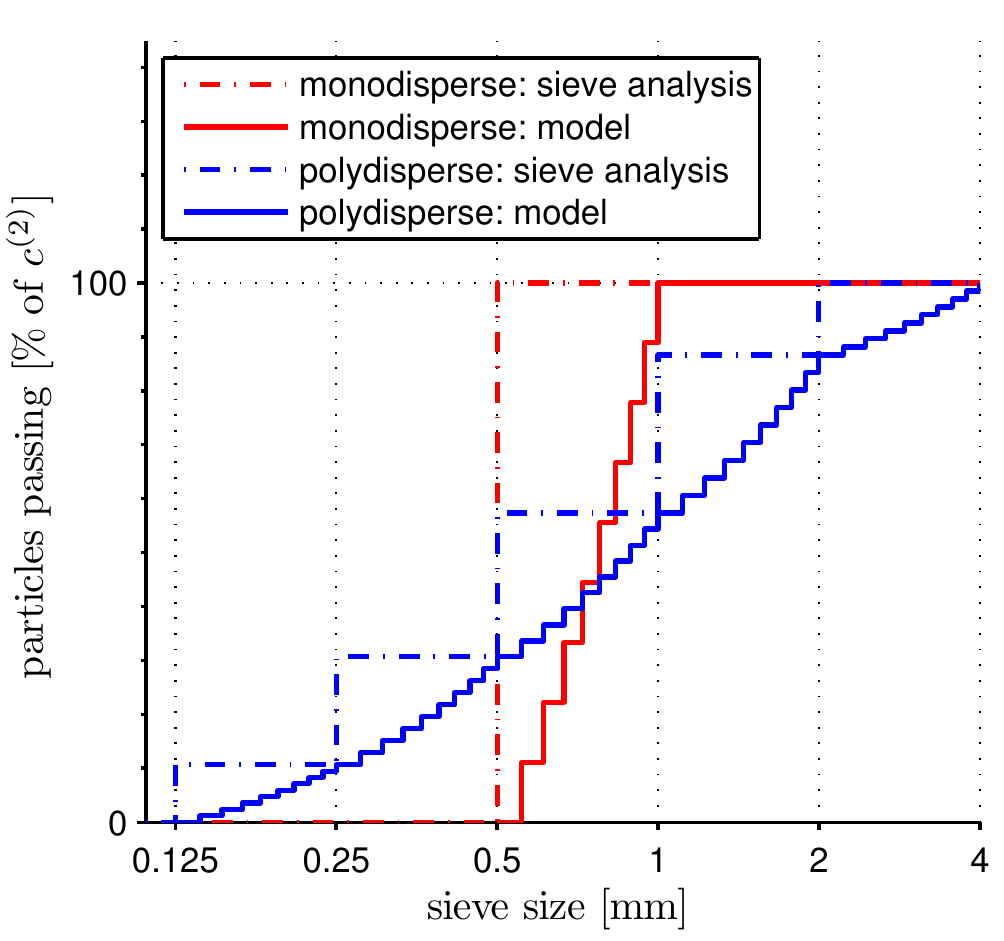}
\caption{Grading curves representing the grain-size distributions of monodisperse and polydisperse aggregates. The sieve analysis provides the distribution into five rather wide intervals, which are further decomposed each into 9 sub-intervals by the linear interpolation for the calculation purposes.}
\label{fig:gradingCurves}
\end{figure}

The model validation was done on two kinds of samples, considering monodisperse and polydisperse particle-size distribution of the aggregates. The term ``monodisperse'' refers to a single to fraction 0.5--1.0~mm, obtained by sieve separation during preparation of the samples. The polydisperse particles were spanning the diameter range from 0.063~mm to 4.0~mm; for illustration see the grading curves plotted in Figure~\ref{fig:gradingCurves}.

\subsection{Elastic Parameters} \label{sec:elastic}

\subsubsection{Matrix Young's Modulus} \label{sec:inputSti}
\begin{figure}[ht]
\centering
   \begin{subfigure}{0.46\linewidth} \centering
     \includegraphics[width=\textwidth]{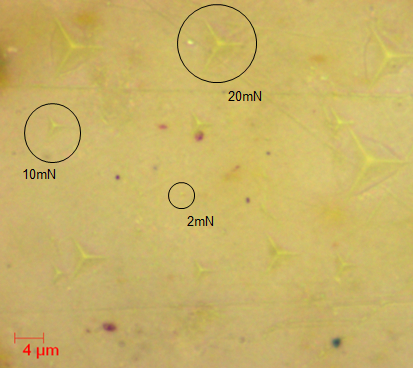}
     \caption{location and size of indents within the lime-metakaolin matrix}
   \end{subfigure}
   \begin{subfigure}{0.46\linewidth} \centering
     \includegraphics[width=\textwidth]{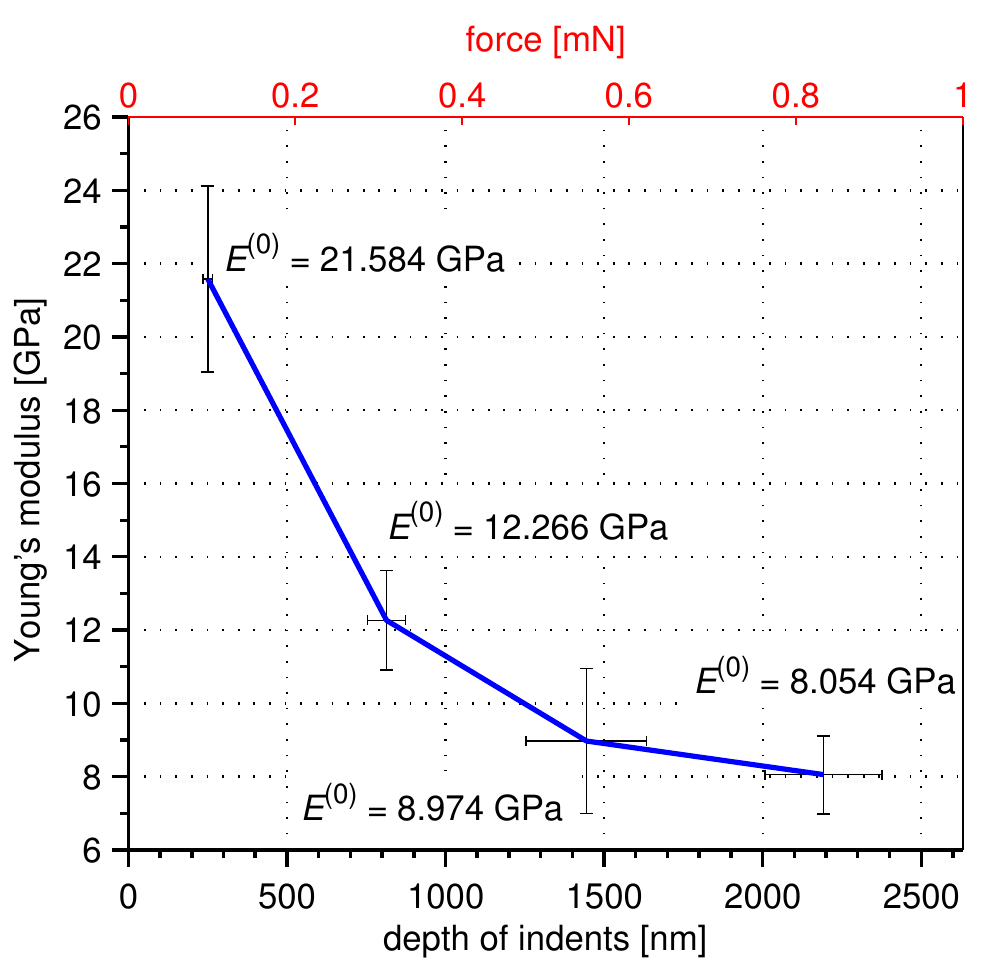}
     \caption{dependence of the effective matrix Young's modulus on the depth of indents; the vertical error bars indicate standard deviations in the measured Young's moduli, horizontal ones represent the deviations in indentation depth}
   \end{subfigure}
\caption{Assessment of the matrix stiffness by means of nanoindentation.}
\label{fig:nanoindentation}
\end{figure}

The matrix elastic stiffness (Young's modulus) was studied by means of quasi-static nanoindentation, the nanohardness tester (CSM Instruments, Switzerland) equipped with Berkovich pyramidal diamond tip. The mortar sample containing sand aggregates was sectioned and polished before the measurement, and a suitable location (away from aggregates) with minimum roughness was selected, Figure~\ref{fig:nanoindentation}. Load-controlled quasi-static indentation test was employed for all imprints, with the load function containing three segments (constant loading at 24~mN/min, 10~seconds holding period, and unloading at 24~mN/min). The indents were evaluated according to the Oliver and Pharr methodology~\cite{OliverPharr_1992}, utilizing the unloading part for the assessment of the material elastic modulus. The holding period was introduced to reduce the creep effects on the elastic unloading~\cite{Nemecek_2009}.

The penetration depth of individual indents varied in order to find the relationship between the penetration depth and the measured Young's modulus. A rapid decrease in elastic stiffness with increasing penetration depth, Figure~\ref{fig:nanoindentation} indicated that the evaluated Young's modulus was affected by the presence of microscale porosity occurring within the indented material volume. According to our measurements, based on mercury intrusion porosimetry, the maximum pore size was established as~900~nm and virtually no pores occurred beyond that limit. As a consequence, the penetration depth about~2,200~nm appeared sufficient to include all nano- and micro-pores, yielding the effective matrix stiffness approximately 8~GPa, see Figure~\ref{fig:nanoindentation}.

Note that a similar study has been performed by Ne\v{z}erka et al.~\cite{Nezerka_2013_pastes} for pure lime-based pastes and their mechanical properties. However, the measured values of Young's modus 3.3~GPa reported therein cannot be used for the modeling of mortars, since the pastes were porous and contained larger voids due to the lack of aggregates. The inclusions contribute to consolidation of the fresh mortar, which results in a much denser matrix.

\subsubsection{Young's Modulus of Aggregates and ITZ}
The Young's modulus of crushed brick fragments was assessed using the resonance method as an average of six measurements on the uncrushed prismatic specimens (40~$\times$~40~$\times$~160~mm). The Young's modulus of river sand was provided by Nilsen and Monteiro~\cite{Nilsen_1993} and the value is in agreement with Daphalapurkar et al.~\cite{Daphalapurkar_2011} who used nanoindentation for the sand elastic stiffness assessment.

According to Yang~\cite{Yang_1998}, the ITZ stiffness reduction in the thickness of 10~$\upmu$m around stiff aggregates is approximately 30~\%, compared to the surrounding matrix. The formation of damaged zone due to shrinkage cracking was also predicted by the numerical model as demonstrated in Figure~\ref{fig:crackFormation}. Therefore, the 30~\% matrix stiffness reduction was also adopted in our micromechanical model. The reinforcement of the interface between the crushed brick fragments and the surrounding matrix by the formation of hydration products~\cite{Baronio_1997_2,Bakolas_2008} was not considered in the model, because their impact on mechanical properties was found to be negligible~\cite{Nezerka_2013_pastes}.

\subsubsection{Poisson's Ratio}
The values of the Poisson's ratio were set according to literature survey. Namely, the value of 0.25 was proposed for the lime-based pastes by Drd\'{a}ck\'{y} and Michoniov\'{a}~\cite{Drdacky_2003}, and we employ the same value also for voids and ITZ. Vorel et al.~\cite{Vorel_2012} considered the value of the Poisson's ratio equal to 0.17 for siliceous sand and the same value was suggested in~\cite{Nezerka_2012_AP} for clay brick fragments.

\subsection{Inelastic Matrix Properties} \label{sec:inelastic}

\subsubsection{Matrix Compressive Strength} \label{sec:inputStrCompress}
The onset of plastic deformation is assumed to take place exclusively in the matrix. Its strength was determined from the destructive uniaxial compression tests carried out on cubic 40~$\times$~40~$\times$~40~mm specimens of lime-metakaolin pastes, as described in~\cite{Nezerka_2013_pastes}.

\subsubsection{Matrix Tensile Strength and Fracture Energy} \label{sec:inputEn}

Since the damage evolution in our model is restricted to the matrix, the tensile strength and fracture energy of other phases was not investigated. Because of a complicated clamping of the samples during the uniaxial tension test and huge scatter of the measured data~\cite{Nezerka_2013_pastes}, the tensile (formally flexural) strength was determined from the three-point bending tests on unnotched simply supported 160~$\times$~40~$\times$~40~mm beams with the distance between supports equal to 120~mm.

The same experimental set-up was employed for the determination of the fracture energy, however, the beams were weakened by a 10~mm notch in the midspan in order to capture the descending part of the load-displacement diagram and avoid snap-back. The fracture energy, $G_\mathrm{f}$, was evaluated directly from the recorded load-displacement diagram using the RILEM approach~\cite{rilem_1985}.

\subsection{Validation Data} \label{sec:validation}

Beside the acquisition of the input data, the purpose of the experimental analysis was to validate the proposed model. To that goal, six mortar specimens representing each batch were cast in prismatic molds 40~$\times$~40~$\times$~160~mm, compacted using shaking table to get rid of excessive air bubbles, and removed from the molds after 24~hours. The amount of water was adjusted so that the fresh mortars fulfilled workability defined by~\v{C}SN~EN~1015-3 and the mortar cone expansion was in the range of~13$\pm$0.5~cm. Curing of mortars was executed at the temperature of 20$\pm$1$^{\: \circ}$C and relative humidity ranging between 65~and~75~\%. The material properties of each mortar mix were assessed after the curing period of 3~months, using the same methods as described in Sections~\ref{sec:inputSti}, \ref{sec:inputStrCompress}, and \ref{sec:inputEn}.

The binder was composed of the same constituents as described in Section~\ref{sec:materials}, i.e. lime and metakaolin in the mass ratio equal to 7:3. The aggregates were either sand or crushed bricks, both of monodisperse and polydisperse particle-size distribution as indicated in Figure~\ref{fig:gradingCurves}. In order to test the model predictions, the samples with polydisperse distribution were prepared in variable binder to aggregate ratio, yielding the aggregate volume fractions equal to $c^{(2)}$=~0.238, 0.384, 0.483, 0.555, and~0.609, respectively.

\section{Results and Discussion} \label{sec:results}
To reproduce the experimental outcomes, the computational procedures described in Section~\ref{sec:model} were employed. The Young's modulus, tensile strength and fracture energy were obtained by an inverse analysis of the stress-strain diagrams predicted by the proposed model\footnote{The MATLAB code {\bf Homogenizator~MT: Composite with Cracks}, can be used for reproduction of results contained in this paper and it is freely available at \url{http://mech.fsv.cvut.cz/~nezerka/software}.}. The simulations of uniaxial compression and tension tests were carried out considering the same composition as pursued when preparing the tested mortars presented in Section~\ref{sec:materials}. The study was mainly focused on the relationship between the effective mortar properties and the volume fraction of aggregates, either crushed bricks or quartz sand, both in mono- and poly-disperse configuration.

\begin{figure}[htp]
\centering
   \begin{subfigure}{0.46\linewidth} \centering
     \includegraphics[width=\textwidth]{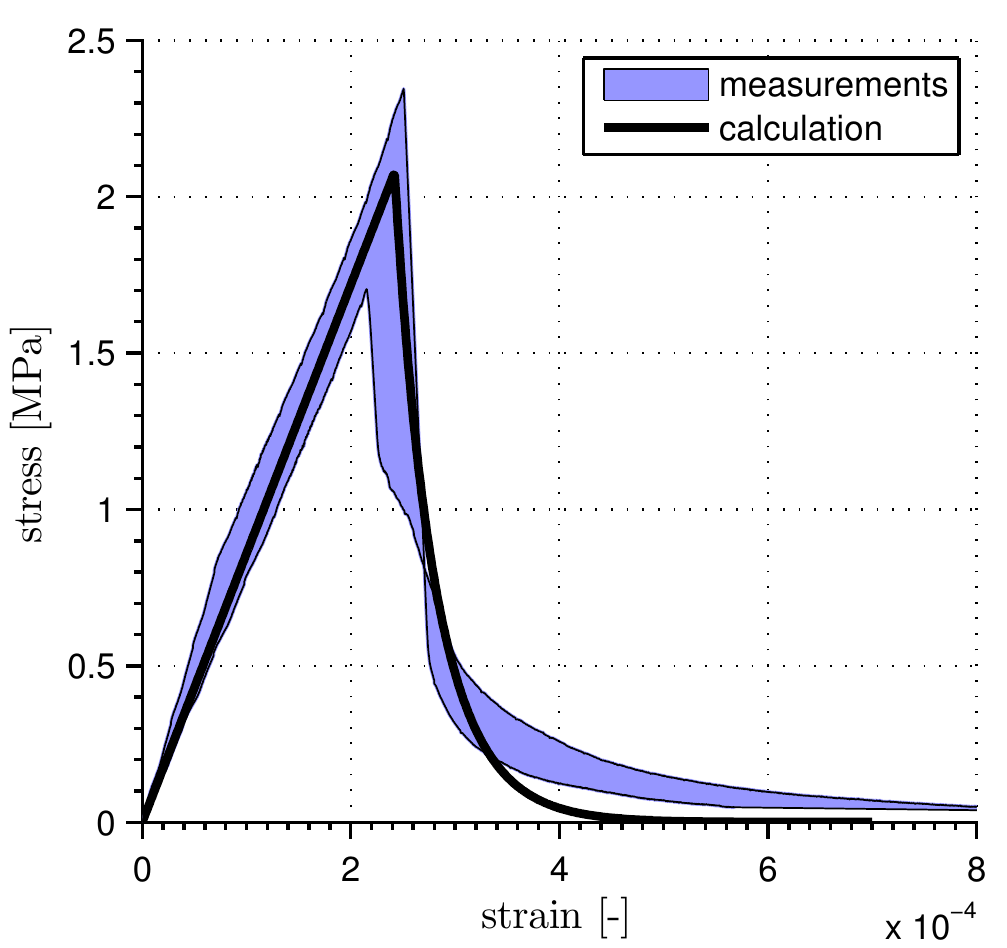}
   \end{subfigure}
   \begin{subfigure}{0.46\linewidth} \centering
     \includegraphics[width=\textwidth]{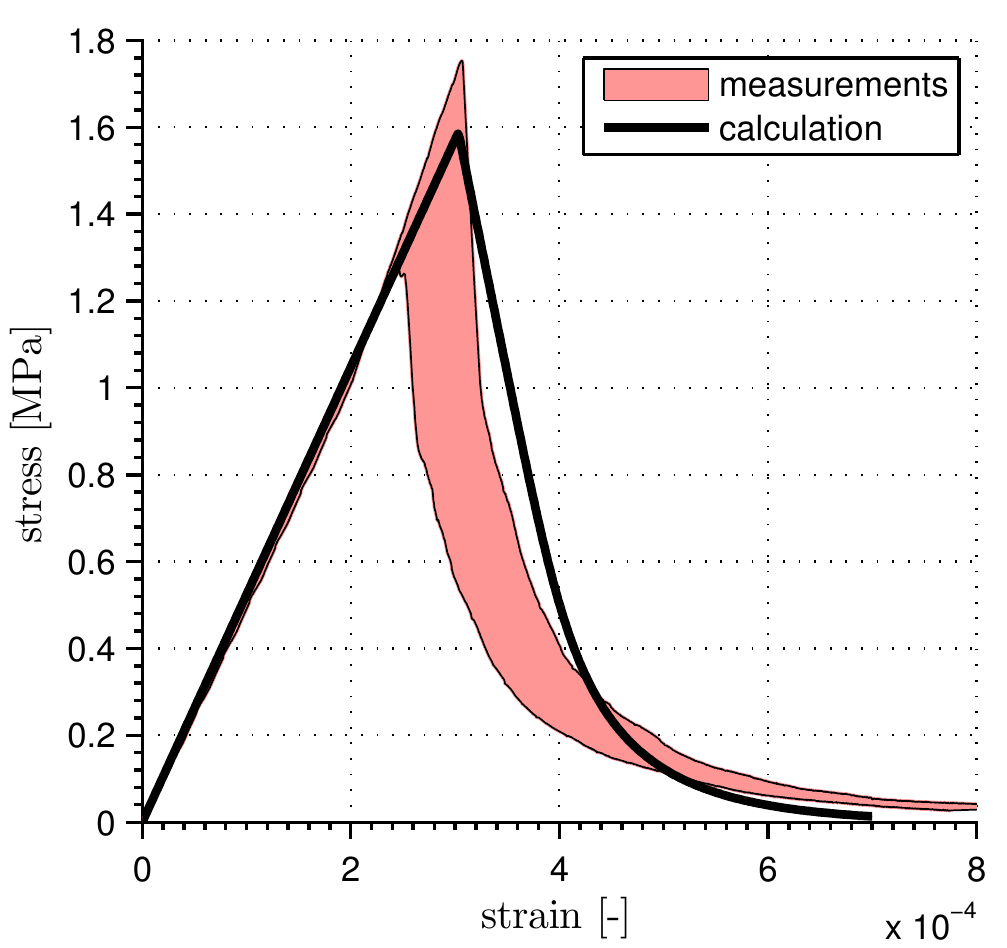}
   \end{subfigure}
\caption{Comparison between the experimentally obtained and calculated stress-strain diagrams in uniaxial tension for the mixes containing monodisperse sand (left) and crushed brick (right) aggregates.}
\label{fig:diagramsComparison}
\end{figure}

The accuracy of our model is demonstrated by means of the uniaxial tensile stress-strain curves in Figure~\ref{fig:diagramsComparison}, provided by the analysis of the samples with ``monodisperse'' particle size distribution, recall Figure~\ref{fig:gradingCurves}. The visual comparison suggests that the model matches well the elastic modulus and the peak strength; the post-peak softening curves are reproduced less accurately, but still provide reasonably accurate values of the fracture energy. 

The purpose of the following Sections~\ref{sec:resultsStiffness}--\ref{sec:resultsEnergy} is to make these observations more quantitative, by investigating the accuracy of the Young's modulus, compressive and tensile strengths, and tensile fracture energy, respectively. After validating the model, the study is closed by determining the optimal sand-to-crushed bricks composition in Section~\ref{sec:mortar_optimization}. We shall use the Pearson correlation coefficient
\begin{align}
\xi
=
\frac{\sum_{i=1}^{n}( x_i - \overline{x} )( y_i - \overline{x})}{%
\sqrt{\sum_{i=1}^{n}( x_i - \overline{x} )^2}
\sqrt{\sum_{i=1}^{n}( y_i - \overline{y} )^2}
}
\end{align}
to quantify the match between the $n$ measured values $x_1, x_2, \ldots, x_n$ and the corresponding model predictions $y_1, y_2, \ldots, y_n$, with e.g. $\overline{x} = \frac{1}{n}\sum_{i=1}^n x_i$.

\subsection{Young's Modulus} \label{sec:resultsStiffness}

\begin{figure}[htp]
\centering
   \begin{subfigure}{0.46\linewidth} \centering
     \includegraphics[width=\textwidth]{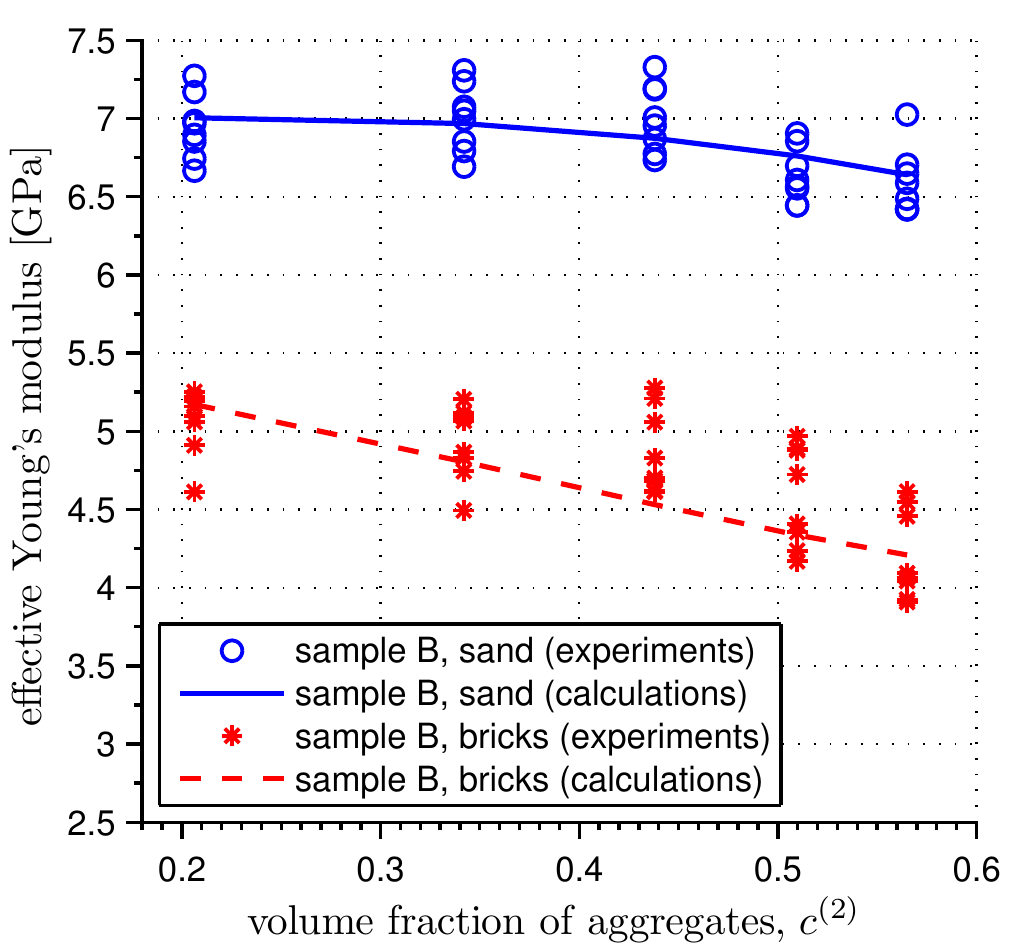}
     \caption{dependence of the effective mortar Young's modulus on the amount of aggregates}
     \label{fig:polydisperseStiffness}
   \end{subfigure}
   \begin{subfigure}{0.46\linewidth} \centering
     \includegraphics[width=\textwidth]{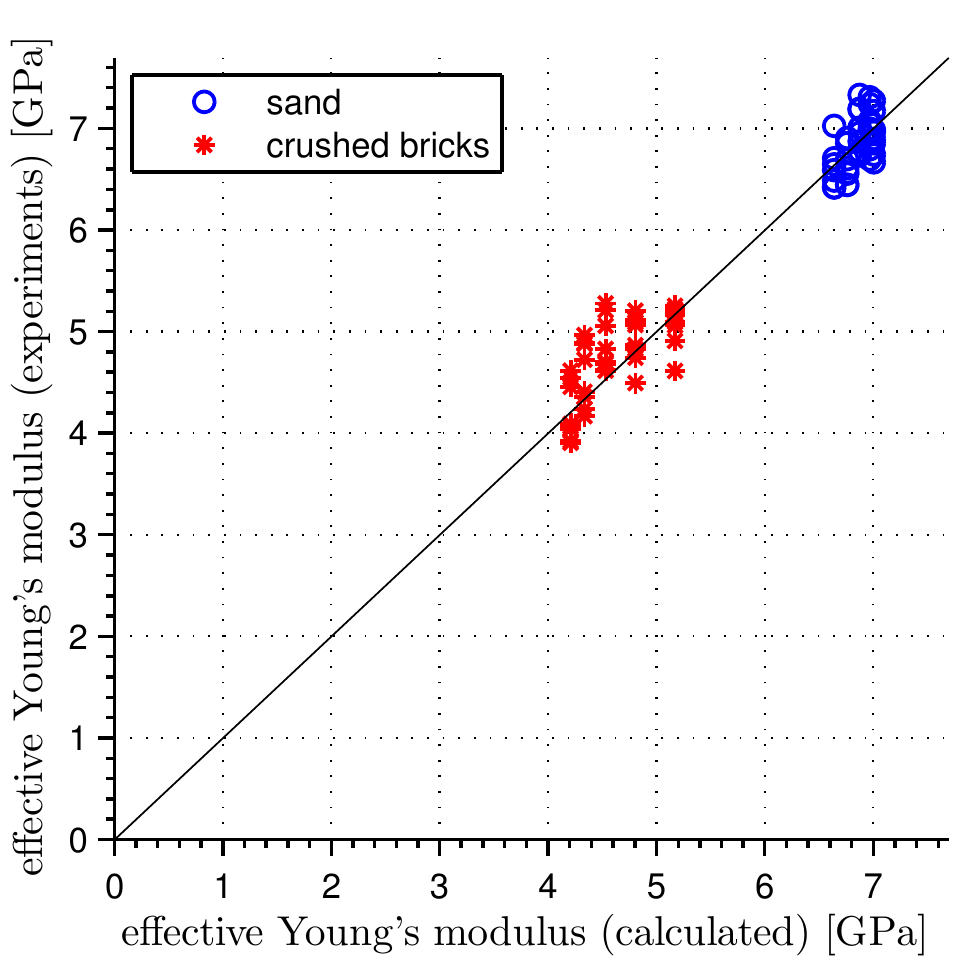}
     \caption{comparison between the calculated and measured values ($\xi$ = 0.973)}
     \label{fig:comparisonStiffness}
   \end{subfigure}
\caption{Comparison between the calculated effective Young's modulus and the experimentally obtained data on mortars containing polydisperse aggregates.}
\label{fig:stiffnessCalculation}
\end{figure}

Figure~\ref{fig:stiffnessCalculation}(a) demonstrates that the Young's modulus of mortars containing sand is constant with the increasing volume fraction of sand particles. This rather surprising behavior is the result of the development of ITZ around the grains, as suggested by Neubauer et el.~\cite{Neubauer_1996}, and the presence of cracks among the shrinkage-constraining grains, observed e.g.~in the study of mortar microstructures by Mosquera et al.~\cite{Mosquera_2006}.

The mortar stiffness reduction with the increasing volume fraction of crushed bricks is a consequence their compliance, which in turn results in the reduction shrinkage-induced cracking and elimination of ITZ formation. Moreover, the hydration promoting nature of the water-retaining crushed brick fragments and their rough surface contribute to a perfect interfacial bond, which has a positive impact on the inelastic properties discussed next.

Given that the values of Pearson correlation coefficient equal 0.973, Figure~\ref{fig:stiffnessCalculation}(b), the model predictions of the effective Young's modulus can be considered very accurate. Such accuracy can be attributed to the introduction of penny-shaped cracks between closely packed aggregates into the homogenization scheme at Level II. Without this step, the method significantly overestimate the stiffness of lime-based mortars, especially for stiff sand aggregates, see~\cite{Nezerka_2014_Brno} for further details.

\subsection{Compressive Strength}

\begin{figure}[htp]
\centering
   \begin{subfigure}{0.46\linewidth} \centering
     \includegraphics[width=\textwidth]{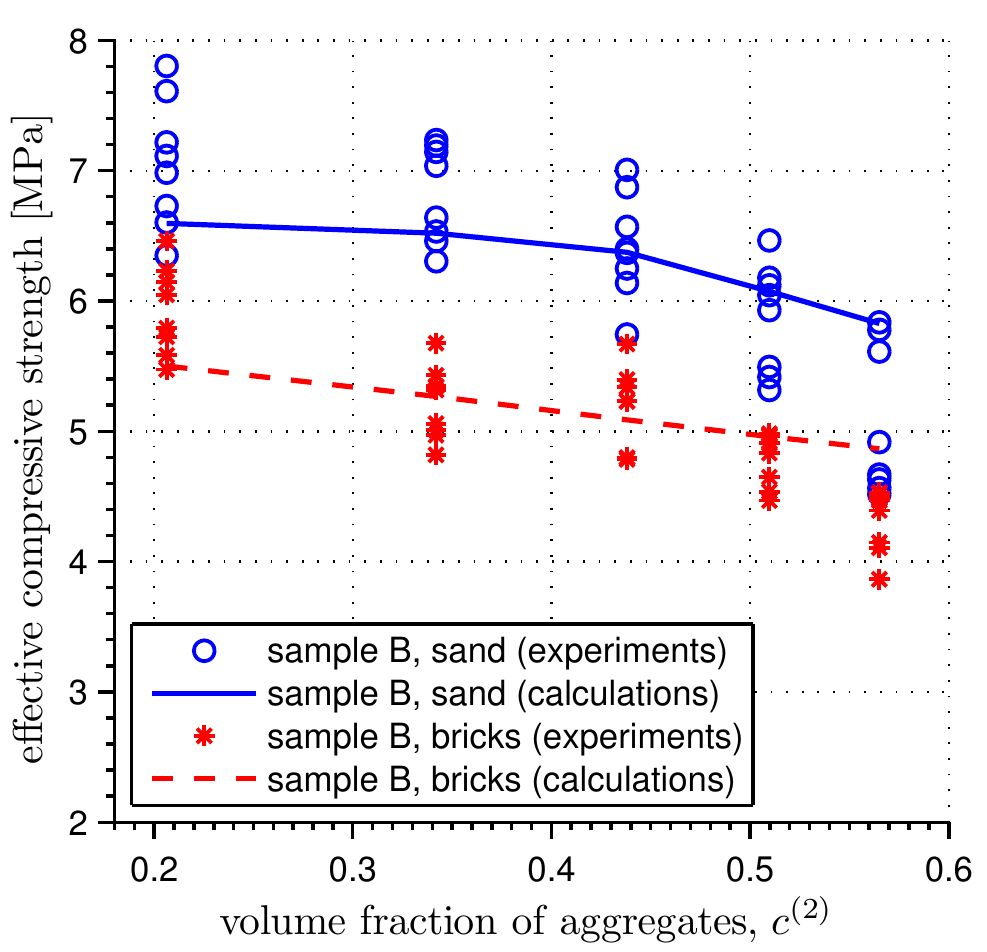}
     \caption{dependence of the effective mortar compressive strength on the amount of aggregates}
     \label{fig:polydisperseCompressiveStrength}
   \end{subfigure}
   \begin{subfigure}{0.46\linewidth} \centering
     \includegraphics[width=\textwidth]{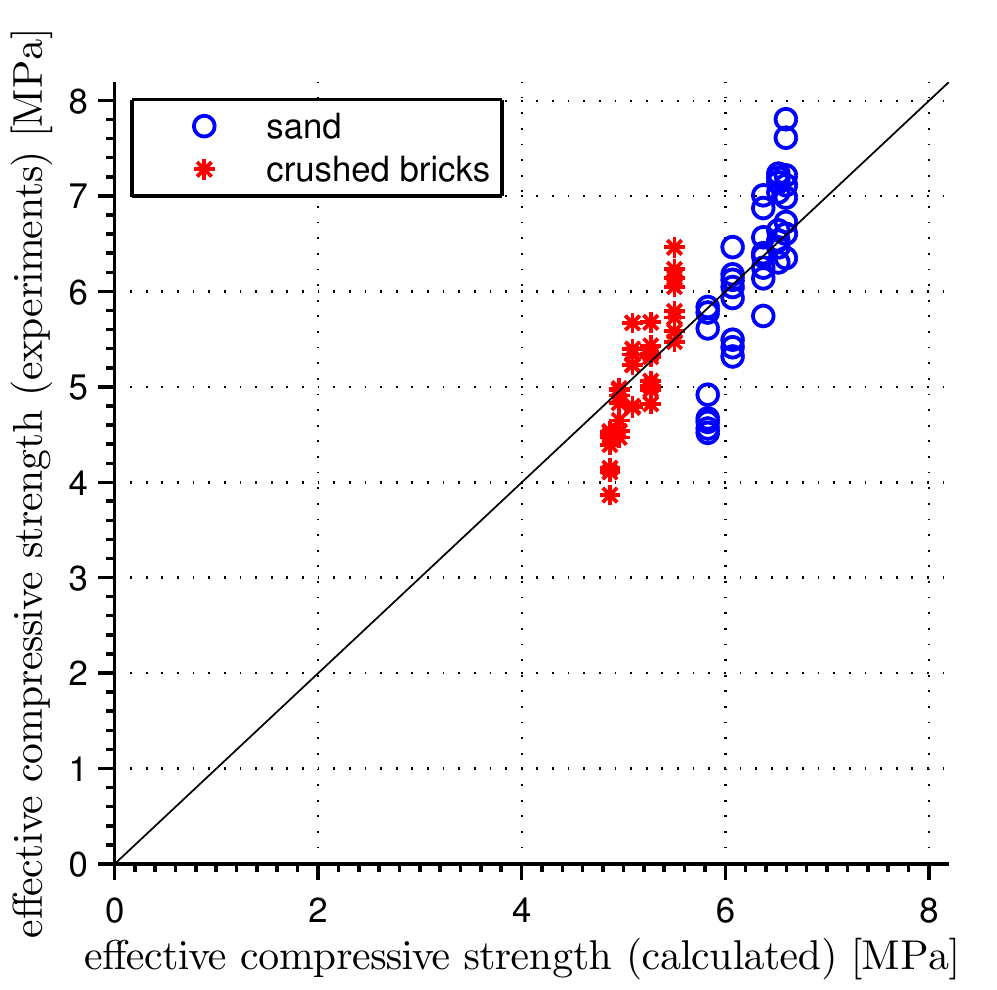}
     \caption{comparison the between the calculated and measured values ($\xi$= 0.846)}
     \label{fig:comparisonCompressiveStrength}
   \end{subfigure}
\caption{Comparison between the calculated effective compressive strength and the experimentally obtained data on mortars containing polydisperse aggregates.}
\label{fig:compressiveStrengthCalculation}
\end{figure}

As follows from our experimental data, Figure~\ref{fig:compressiveStrengthCalculation}(a), and from independent findings by Lanas et al.~\cite{Lanas_2004}, the compressive strength of mortars containing sand grains should be higher than of those with crushed brick fragments and its value should decrease with increasing aggregate volume fractions, because of the stress concentration in the matrix phase. Both trends are correctly reproduced by our model both quantitatively and qualitatively; as visible from Figure~\ref{fig:compressiveStrengthCalculation}(b), the agreement between the model predictions and experiments are of a similar accuracy as for the Young's modulus.

\subsection{Tensile Strength}

\begin{figure}[htp]
\centering
   \begin{subfigure}{0.46\linewidth} \centering
     \includegraphics[width=\textwidth]{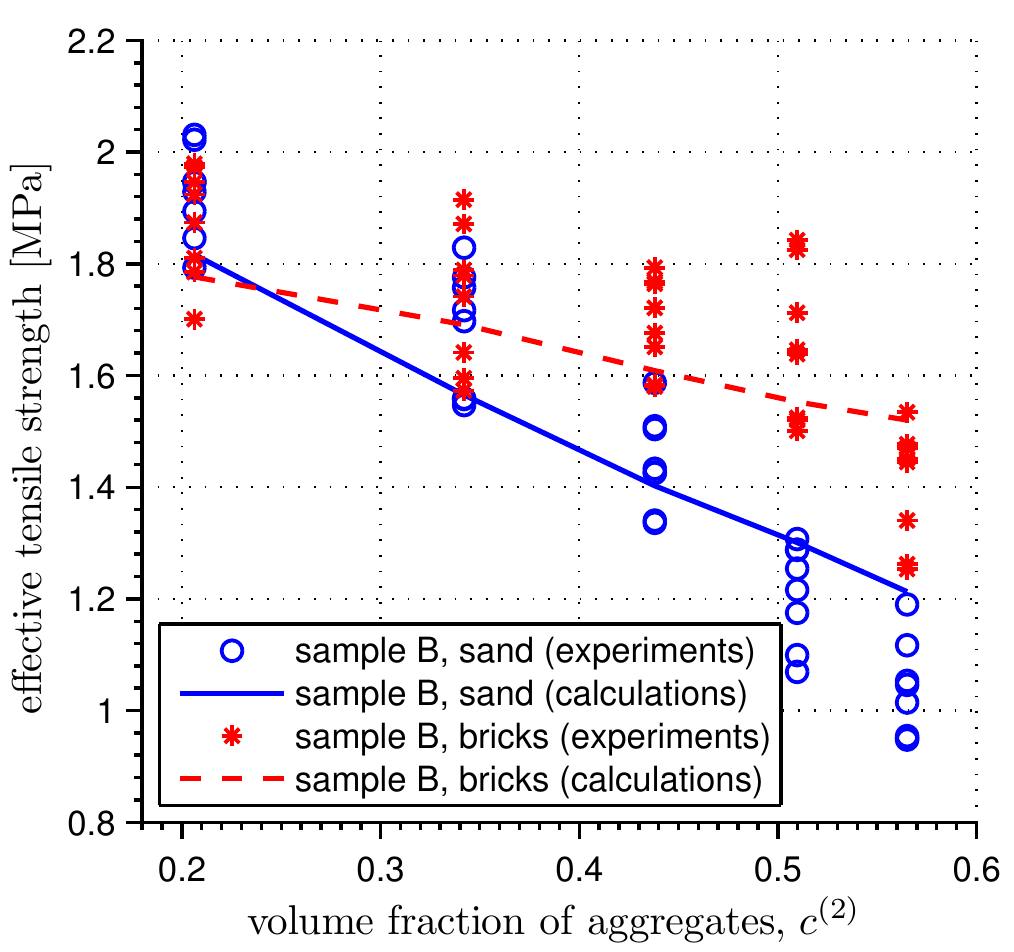}
     \caption{dependence of the effective mortar tensile strength on the amount of aggregates}
     \label{fig:polydisperseStrength}
   \end{subfigure}
   \begin{subfigure}{0.46\linewidth} \centering
     \includegraphics[width=\textwidth]{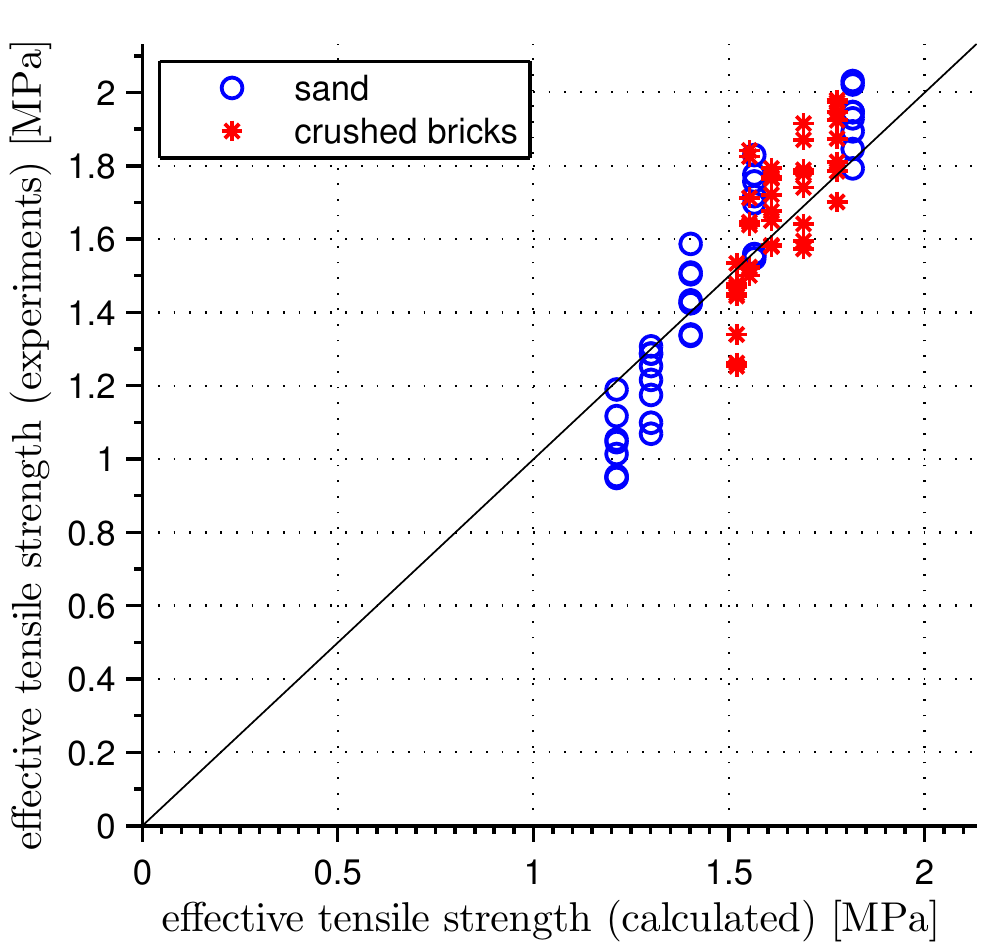}
     \caption{comparison the between the calculated and measured values ($\xi$ = 0.916)}
     \label{fig:comparisonStrength}
   \end{subfigure}
\caption{Comparison between the calculated effective tensile strength and the experimentally obtained data on mortars containing polydisperse aggregates.}
\label{fig:strengthCalculation}
\end{figure}

According to experimental measurements, Figure~\ref{fig:strengthCalculation}(a), the tensile strength of lime-based mortars is also reduced with the increasing volume fraction of aggregates, which is also in agreement with the findings of Lanas et al.~\cite{Lanas_2004}. The tensile strength reduction with the increasing amount of aggregates is more pronounced in the case of mortars containing sand. This phenomenon is reflected by higher strain concentrations in the matrix, responsible for the onset of damage at lower levels of externally applied macroscopic strain. The agreement between the model and experiments for the tensile strength is slightly worse that in the case of elastic stiffness, compare Figure~\ref{fig:strengthCalculation}(b) to Figure~\ref{fig:stiffnessCalculation}(b), but we consider such accuracy to be sufficient for engineering purposes, especially when taking into account the scatter of experimental data, Figure~\ref{fig:strengthCalculation}(a).

\subsection{Fracture Energy}\label{sec:resultsEnergy}

\begin{figure}[htp]
\centering
   \begin{subfigure}{0.46\linewidth} \centering
     \includegraphics[width=\textwidth]{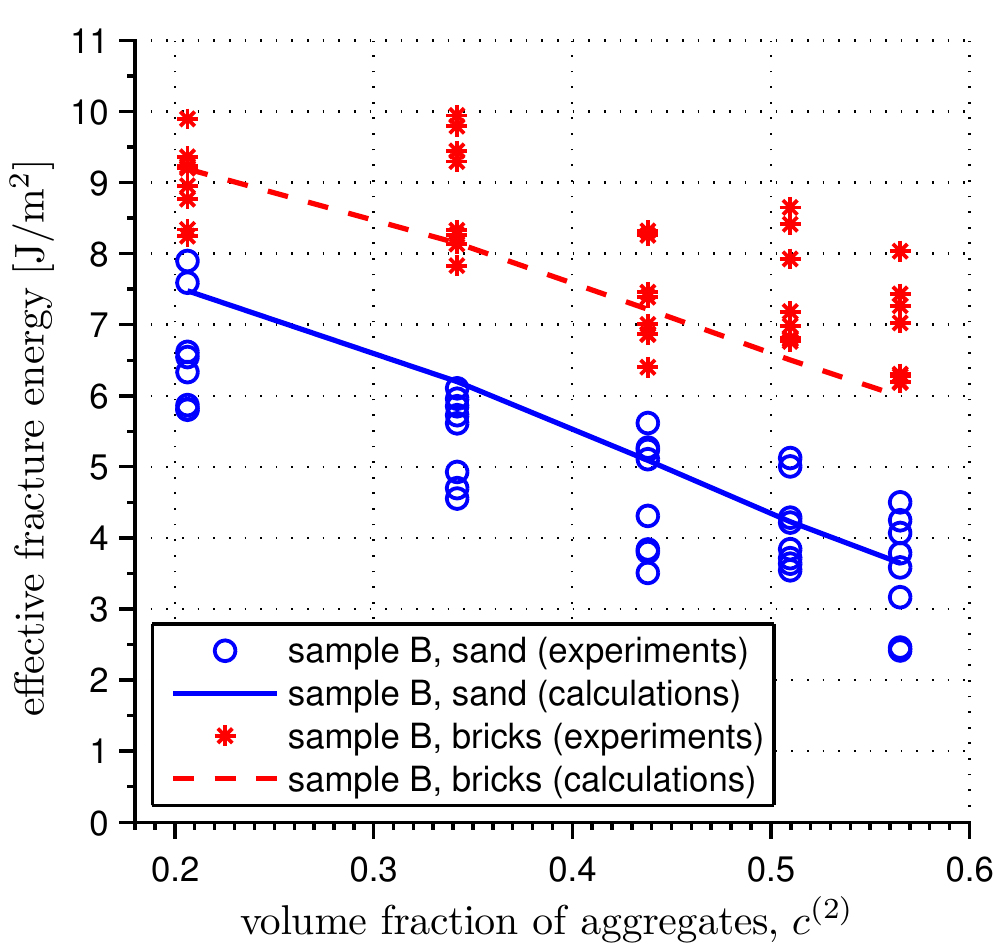}
     \caption{dependence of effective mortar fracture energy on the amount of aggregates}
     \label{fig:polydisperseEnergy}
   \end{subfigure}
   \begin{subfigure}{0.46\linewidth} \centering
     \includegraphics[width=\textwidth]{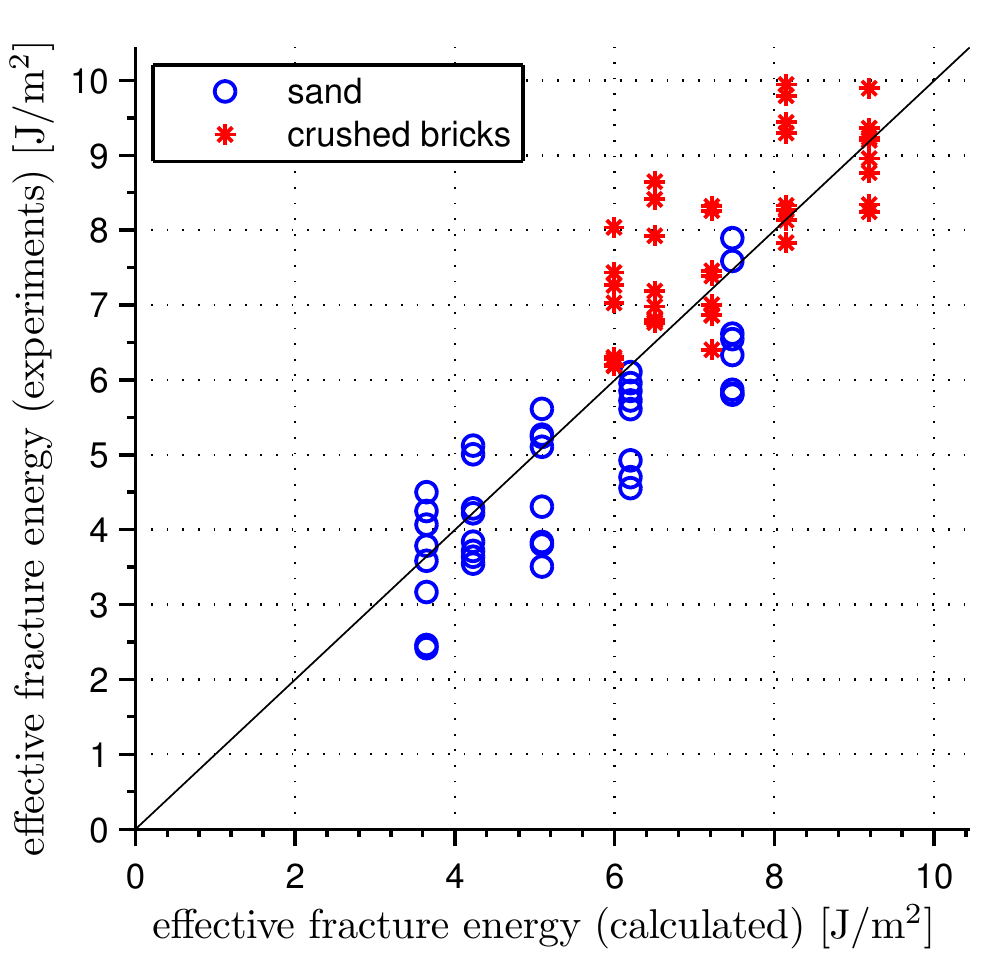}
     \caption{comparison the between the calculated and measured values ($\xi$ = 0.879)}
     \label{fig:comparisonEnergy}
   \end{subfigure}
\caption{Comparison between the calculated effective fracture energy and the experimentally obtained data on mortars containing polydisperse aggregates.}
\label{fig:energyCalculation}
\end{figure}

According to both, model and experiments, the values of fracture energy were higher in the case of mortars containing crushed brick aggregates, also because of the lower contrast in elastic properties leading to the decrease in stress concentrations. In consequence, these mortars accumulate more energy by allowing to reach higher values of elastic deformation, Figure~\ref{fig:diagramsComparison}. Again, the predictions of our model are satisfactory considering the relatively high scatter of the experimentally obtained data, as indicated by $\xi$ equal to 0.879, see Figure~\ref{fig:energyCalculation}(b).

In summary, our experimental and model-based results confirm that with respect to sand-reinforced lime mortars, the mortars containing crushed brick particles provide lower stiffness, higher tensile and compressive strengths, and fracture energies. These factors combined allow to achieve significantly larger maximum strain and ductility in tension, which probably contributes to the increased earthquake resistance. Such conclusions correspond well with the findings by e.g. Moropoulou et al. or Baronio et al.~\cite{Moropoulou_2000_2, Baronio_1997}, but our model provides a different explanation for the emergent behavior. Specifically, the improved mechanical performance is in~\cite{Moropoulou_2000_2} primarily attributed to the formation of hydration products at the interface between the lime matrix and fragments of crushed bricks. Even though this phenomenon has been recently confirmed by nano-indentation~\cite{Nezerka_2014_nanoindentation}, our model reveals that it is of secondary importance. The dominant mechanism is found in the reduction of the shrinkage cracks due to more compliant cricks at Level II of the model; recall Figure~\ref{fig:criticalRatio} and see~\cite{Nezerka_2014_Brno} for an explicit demonstration in the elastic regime.

\subsection{Mortar Mix Optimization}\label{sec:mortar_optimization}

\begin{figure}[htp]
\centering
   \begin{subfigure}{0.46\linewidth} \centering
     \includegraphics[width=\textwidth]{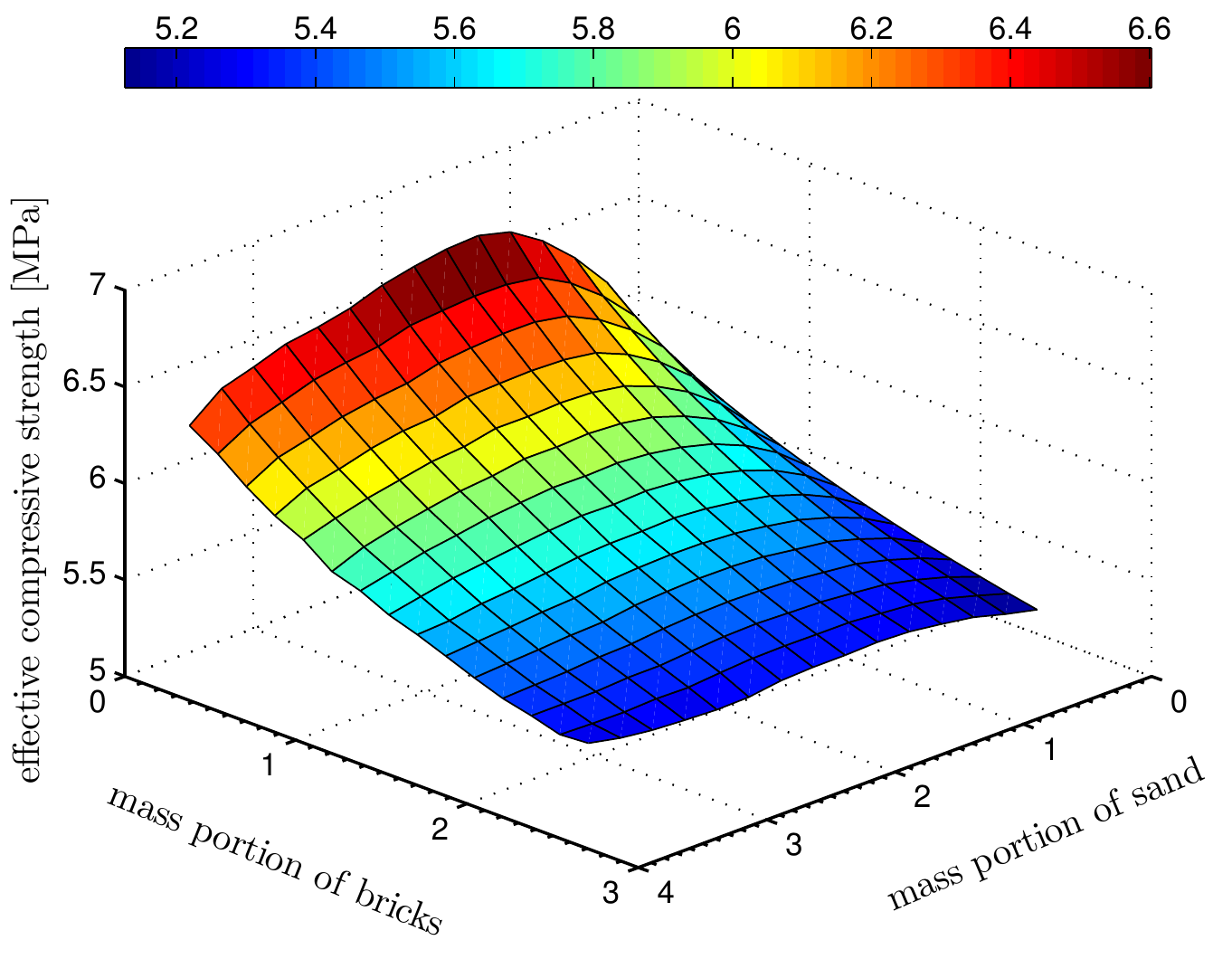}
   \end{subfigure}
   \begin{subfigure}{0.46\linewidth} \centering
     \includegraphics[width=\textwidth]{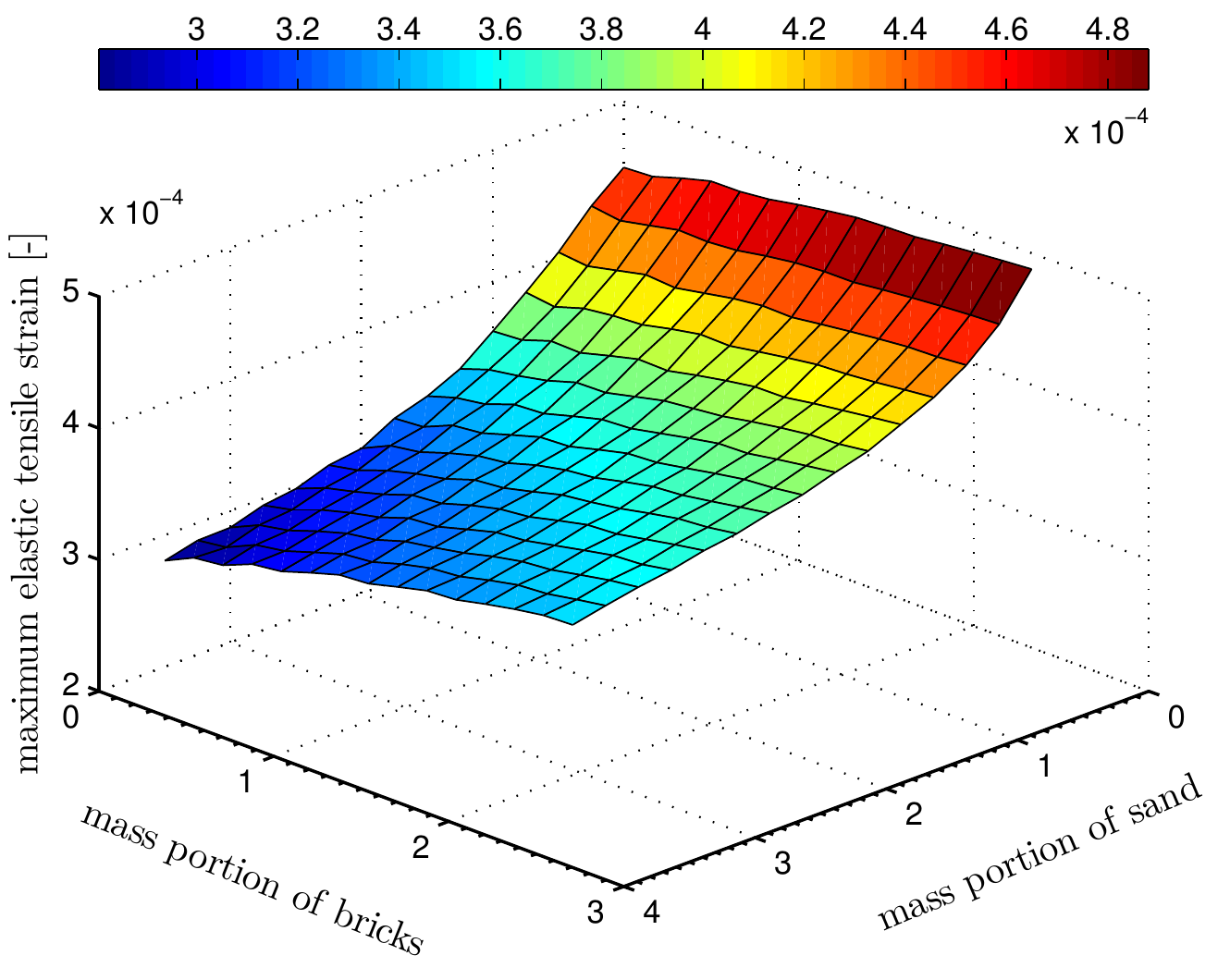}
   \end{subfigure}
\caption{Dependence of the mortar compressive strength (left) and maximum elastic deformation in tension (right) on the amount of sand and crushed brick fragments, considering polydisperse aggregates.}
\label{fig:mixOptimization}
\end{figure}

Having validated the model, we now proceed with its application to the design of the optimal ratio between the amount of sand and crushed brick fragments to achieve maximum elastic deformation in tension and compressive strength.
Figure~\ref{fig:mixOptimization} shows the predicted dependence of the target effective mortar properties, from which we suggest the optimum binder~/~sand~/~crushed bricks ratio equal to 1:1:1.5. These values are close to the generally established minimum binder~/~sand mass ratio 1:3~\cite{Wilk_2013} in order to avoid excessive cracking, especially when recalculated to volumetric fractions due to low mass density of crushed brick aggregates. Based on our experience, such ratio also provides optimum workability when used as a bed joint masonry mortar.

\section{Conclusions}\label{sec:conclusions}
The simple micromechanics-based model of lime-based mortars for the estimation of their elastic stiffness, compressive and tensile strength, and fracture energy was proposed and validated against experiments. The model consists of two levels, where the lower level describes the interaction among individual components of a mortar mix, while the upper scale accounts for the shrinkage-induced cracks that significantly influence the overall mechanical performance. As for the prediction of the effective parameters, the Mori-Tanaka / Dilute Approximation was used to estimate the overall stiffness, a $J_2$-based failure condition involving the average stress in matrix is adopted under compression, and under tension we employed the incremental Mori-Tanaka method coupled with the isotropic damage law and crack bend theory.

Based on the presented results, we have found that the model correctly predicts
\begin{enumerate}\itemsep=0pt
  \item the elastic stiffness of mortars containing sand, which does not increase with the increasing volume fraction of the aggregate due to the formation of ITZ and shrinkage-induced cracks between closely packed grains,
  \item the elastic stiffness of mortars containing crushed brick fragments, which is also reduced by the addition of crushed bricks as a consequence of their low stiffness, and due to formation of shrinkage-induced cracks between the particles, however in lesser extent compared to the stiff sand grains,
  \item the mortar compressive strength, which is higher in the case of mortars containing stiff sand grains, and decreases with the increasing amount of the aggregates,
  \item the mortar tensile strength, significantly reduced by an increase of the amount of aggregates; the effect is more pronounced in the case of sand grains, rather than compliant crushed brick fragments,
  \item the mortar fracture energy, being higher if the crushed brick fragments replace sand grains, since the more compliant mortars can reach higher elastic deformation.

\end{enumerate}
For all these quantities, we reached both the quantitative and qualitative agreement between the experimental results and the model predictions.

The validated model can be used for obtaining the input data for numerical modeling of mortars at meso- and macroscale, or to find an optimal composition of the mix with respect to structural performance in masonry structures. As for the latter application, we invite the interested reader to our follow-up paper~\cite{Nezerka_2015-piers}, where we confirm by full-scale testing that the optimized mortar delivers superior mechanical performance and durability with respect to conventional lime-based mortars with sand reinforcement.

\section*{Acknowledgments}
The authors would like to thank for the financial support provided by the Ministry of Culture of the Czech Republic under the project NAKI, No.~DF11P01OVV008, and for the support by the Czech Science Foundation, project No.~GA13-15175S.

\end{document}